\documentclass[acmtog,nonacm]{acmart} 

\usepackage{booktabs} 

\acmPrice{15.00}

\setcopyright{rightsretained}
\acmJournal{TOG}
\acmYear{2024}
\acmVolume{00}
\acmNumber{00}
\acmArticle{00}
\acmMonth{00}
\acmDOI{10.1145/00...}

\setcitestyle{square}

\newif\ifmarkups
\markupsfalse

\newif\ifsupplement
\supplementtrue

\usepackage[T1]{fontenc}

\usepackage{microtype}

\usepackage{amsmath}
\newcommand{\eqn}{ 
  \refstepcounter{equation}
  \eqno \hbox{{\normalfont \normalcolor (\theequation) }}
}
\newcommand{\mbf}[1]{\boldsymbol{\mathbf{#1}}} 

\usepackage{algorithm}
\usepackage[noend]{algorithmic}
\usepackage{matlab-prettifier} 
\definecolor{darkgreen}{RGB}{34,139,34}

\usepackage[utf8]{inputenc}
\usepackage{alphabeta}
\usepackage{textalpha}

\usepackage{fontawesome}

\usepackage{gensymb}
\DeclareUnicodeCharacter{00B0}{\degree}

\usepackage{graphicx} 
\usepackage{subcaption} 
\usepackage{wrapfig} 
\usepackage{booktabs} 
\usepackage{enumitem} 
\usepackage{xcolor}
\colorlet{RED}{red} 

\graphicspath{{img/}}

\newcommand{\user}[2]{\textsf{#2}} 

\ifmarkups
  \usepackage{pdfcomment} 
  \newcommand{\rev}[2]{{\color{red}{#1}}}
  \newcommand{\todo}[1]{{\color{blue}ToDo: *** {#1} ***}}
  \newcommand{\pen}[1]{{\textbf{\color{brown}[P.EN: #1]}}}
  
  \newcommand{\cmtdh}[1]{{{\color{gray}[DH: #1]}}}
  \usepackage[nolabel]{showlabels,rotating}
  
\else
  \usepackage[final]{pdfcomment} 
  \newcommand{\todo}[1]{}
  \newcommand{\rev}[2]{{#1}}
  \newcommand{\pen}[1]{}
  
  \newcommand{\cmtdh}[1]{}
\fi

\begin{document}


\title[View-Independent Adjoint Light Tracing for Lighting Design Optimization]{View-Independent Adjoint Light Tracing \\for Lighting Design Optimization}

\author{Lukas Lipp}
\email{{lukas.lipp@cg.tuwien.ac.at}}
\author{David Hahn}
\author{Pierre Ecormier-Nocca}
\affiliation{%
  \department{Rendering and Modeling Group}
  \institution{TU Wien}
}

\author{Florian Rist}
\affiliation{%
  \department{Department for Design}
  \institution{TU Wien}
}
\affiliation{%
  \department{Applied Mathematics and Computational Sciences}
  \institution{KAUST}
}

\author{Michael Wimmer}
\affiliation{%
  \department{Rendering and Modeling Group}
  \institution{TU Wien}
}

\renewcommand{\shortauthors}{Lipp, Hahn, Ecormier-Nocca, Rist, and Wimmer}

\begin{abstract}

\section*{Abstract}

Differentiable rendering methods promise the ability to optimize various parameters of 3d scenes to achieve a desired result.
However, lighting design has so far received little attention in this field.
In this paper, we introduce a method that enables continuous optimization of the arrangement of luminaires in a 3d scene via differentiable light tracing.
%
Our experiments show two major issues when attempting to apply existing methods from differentiable path tracing to this problem:
first, many rendering methods produce images, which restricts the ability of a designer to define lighting objectives to image space.
Second, most previous methods are designed for scene geometry or material optimization and have not been extensively tested for the case of optimizing light sources.
Currently available differentiable ray-tracing methods do not provide satisfactory performance, even on fairly basic test cases in our experience.
%
In this paper, we propose a novel adjoint light tracing method that overcomes these challenges and enables gradient-based lighting design optimization in a view-independent (camera-free) way.
Thus, we allow the user to paint illumination targets directly onto the 3d scene or use existing baked illumination data (e.g.,~light maps).
Using modern ray-tracing hardware, we achieve interactive performance.
We find light tracing advantageous over path tracing in this setting, as it naturally handles irregular geometry, resulting in less noise and improved optimization convergence.
%
We compare our adjoint gradients to state-of-the-art image-based differentiable rendering methods.
We also demonstrate that our gradient data works with various common optimization algorithms, providing good convergence behaviour.
Qualitative comparisons with real-world scenes underline the practical applicability of our method.

Peer-reviewed version: \url{https://dl.acm.org/doi/10.1145/3662180}


\end{abstract}



\begin{teaserfigure}
\centering
\includegraphics[width=\columnwidth]{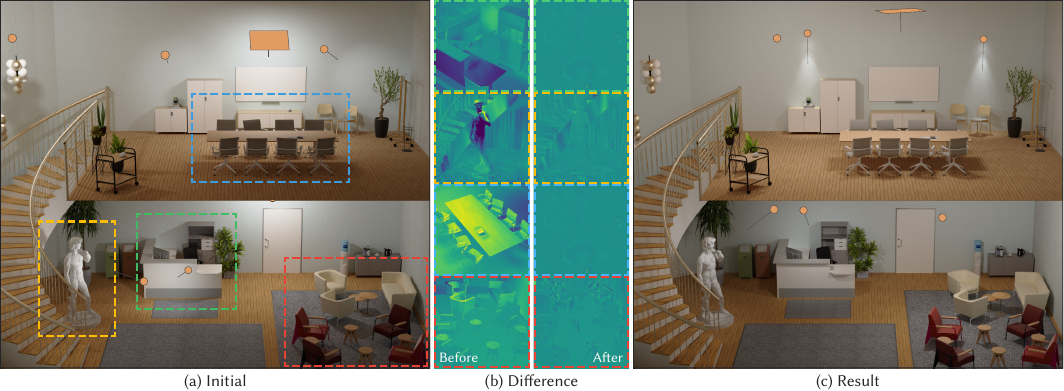}
\caption{
Optimizing lighting design parameters for a large office scene with our view-independent differentiable light-tracing method:
starting from an initial lighting setup (a) and a given ground-truth illumination target on the scene geometry, we perform gradient-based optimization, which closely recovers the ground truth in our solution (c).
The close-up views (b) show the difference to the target before and after optimization (left and right columns respectively).
} \vspace{0.5cm}
\label{fg:large-office}
\end{teaserfigure}

\maketitle

\vfill\eject 

\section{Introduction}

Lighting plays an often overlooked but central role in our daily lives. It contributes to making a home feel relaxing, a workplace efficient, and is subject to important requirements in professional environments such as office spaces, hospitals, and factories.
While artistic applications usually require manual luminaire placement, simple grid-like patterns are widespread in office spaces. Both scenarios can benefit from automated optimization, reducing the workload on artists, or producing more comfortable office lighting.
Widely used commercial lighting design tools, such as DIALux \shortcite{Dialux2022} and Relux~\shortcite{Relux2022}, are limited to simple light interactions and require manual editing of luminaires to achieve the desired result.

Physically based \emph{primal} rendering, i.e.,~simulating light transport in a 3d scene according to the rendering equation, has a long-standing history
in computer graphics.
Conversely, \emph{inverse} rendering methods aim to match a scene's appearance to one or more target images.
These existing \emph{view-dependent} methods are technically capable of optimizing light sources in an \emph{image-based} framework, but have practical limitations: as indicated by our comparisons, their convergence behaviour and runtime performance are insufficient for interactive lighting optimization.
Furthermore, they quickly become problematic for complex scenes, as many cameras are required to capture all areas of interest.
Additionally, this approach would raise the question of how many (and where) cameras should be placed, and how to define their illumination targets.
Consequently, we avoid cameras in favour of a view-independent approach.

Currently, no solution for \emph{view-independent} (i.e., camera-free) differentiable rendering exists.
Moving from camera-based inverse renderers (Mitsuba, PSDR) to view-independent optimization requires a spatio-directional data structure that can be updated and edited quickly.
In order to achieve interactive performance, we also require an efficient and robust gradient formulation that can be evaluated on the GPU.
In this paper, we propose an interactive and fully view-indepen\-dent inverse rendering framework, building upon a novel analytical adjoint formulation.
The absence of a predefined camera view, and optimization parameters that are associated with light sources, motivate a light-tracing global-illumination method, instead of more common path-tracing approaches.

Previously, automatic differentiation (AD) has been widely used.
Our tests show that applying AD directly to light tracing suffers from systematic errors that prevent optimization convergence.
Zhang et al.~\shortcite{Zhang2020} describe an AD path-space formulation, which is, in principle, also applicable to paths constructed via light tracing.
Conversely, we present an \emph{analytically differentiable adjoint light-tracing} method, which enables a GPU-accelerated implementation.

Overall, we present the following contributions in this paper:
\begin{itemize}
\item An efficient adjoint gradient formulation for differentiable light tracing, which outperforms comparable path-tracing approaches, and
\item an efficiently updatable view-independent radiance data structure.
\item This combination allows us to solve lighting design optimization tasks, while taking global illumination into account.
\item We also present a novel visualization of point-wise adjoint gradient contributions.
\end{itemize}

\section{Related Work}

Here, we briefly summarize relevant work on global-illumination rendering, before discussing recent advances in differentiable rendering.
Furthermore, we introduce approaches that have investigated the lighting design problem from other directions, such as procedural modelling. 

\subsection{Primal rendering}
In the following, we classify related work in four categories, as illustrated in the inset image. 
Path tracing, one of the most fundamental methods for physically based rendering, applies Monte Carlo integration to solve the rendering equation \cite{Kajiya1986,Veach1998}.
\rev{}{The main idea is to render an image of a virtual scene by following many, randomly sampled, paths starting from a virtual camera and accumulating colour contributions from around the scene at each pixel of the image.}
Similar methods have been extensively used to render physically realistic images (category a). Pre-computed light-transport methods, on the other hand, such as radiosity \cite{Goral1984,Greenberg1986}, focus on solving global illumination for all surfaces in a 3d scene (category~b) rather than for a view-dependent image.
\begin{wrapfigure}{r}{0.4\columnwidth}
  \vspace{-0.3cm}
  \hspace{-0.6cm}
  \includegraphics[width=0.44 \columnwidth]{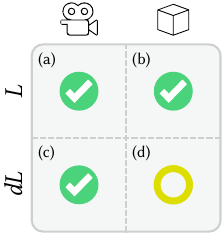}
  \vspace{-0.2cm}
\end{wrapfigure} 
More recently, these radiosity methods have been extended to incorporate glossy surfaces \cite{Sillion1991,Sloan2002,Krivanek2005,Hasan2009,Omidvar2015} and also enable fast incremental updates allowing for interactive frame rates \cite{Luksch2019}.
Similarly, photon mapping \cite{Jensen1995} is a two-pass process, where illumination data is first cast into the 3d scene by tracing light paths, while the second pass gathers this data to compute the final image.
We take inspiration from these latter approaches, constructing a spatio-directional radiance data structure in order to enable \emph{view-independent} differentiable rendering (category d).
The key ideas of our adjoint light-tracing approach would in principle also be applicable to other pre-computed light-transport methods and radiance-field data structures, such as the recent work by Yu et al.~\shortcite{Yu2021}, as well as photon tracing, virtual point light, or light probe approaches.

\subsection{Differentiable rendering}
Previous work in (physically based) \emph{inverse} rendering has, so far, focused on camera-based methods (category c), where the goal is to optimize scene parameters (e.g., materials, textures, or geometry)~\cite{Gkioulekas2013,Khungurn2015,Gkioulekas2016,Liu2018,Li2018,Zhang2019,Nimier-David2019,Nimier-David2020,Zeltner2021}.
These methods define an optimization objective on the difference between a rendered and a target image.
A strong focus has been on optimizing textures, while only few methods have considered optimizing light sources, which affect the result much more globally throughout the scene.
Nimier-David et al.~\shortcite{NimierDavid2021} describe a texture-space sampling strategy to improve efficiency for multi-camera setups, while our approach removes cameras altogether.

Pioneering works in differentiable path tracing \cite{Nimier-David2019,Loubet2019} have relied on code-level automatic differentiation (AD), recording a large computation graph, which is then traversed in reverse to compute the objective function gradient.
However, AD fails for standard light tracing (Alg.~\ref{alg:fwd}; also known as particle tracing), because each ray carries a constant amount of radiative flux (power) along the ray, independently of its direction or distance travelled. Consequently, parameters that affect these quantities (but not the per-ray flux) cannot be correctly differentiated with AD.
The PSDR approach \cite{Zhang2020} side-steps this issue by evaluating radiative transport in a path-space formulation before differentiation.
One of our key insights is that we can take inspiration from their ``material-form differential path integral'' and construct a differentiation of light tracing itself for lighting parameters.
Instead of using automatic differentiation, we formulate an analytical adjoint state per ray, which allows for an efficient GPU-based implementation.

The basic idea of adjoint methods\rev{}{ (also known as backpropagation in machine learning)} is to reverse the flow of information through a simulation to find objective function gradients with respect to optimization parameters \cite{Bradley2019,Geilinger2020,Stam2020}.
For path tracing, Mitsuba~3 \cite{Jakob2022} provides a similar adjoint method, called ``radiative backpropagation'' \cite{Nimier-David2020}, which avoids the prohibitive memory cost of AD. They extend this approach to the reparameterized ``path-replay backpropagation'' integrator \cite{Vicini2021}.

\begin{figure}[t]
\includegraphics[width=\columnwidth]{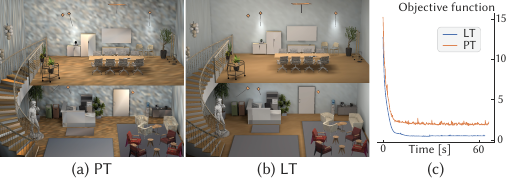}
\caption{
Comparing light tracing to path tracing on a large scene using our radiance data structure: While path tracing (a) would need more advanced sampling strategies to deal with a large non-uniform mesh, light tracing (b) naturally focuses samples on brighter areas, leading to a less noisy result and therefore improved optimization convergence (c) at equal runtime.
}
\label{fg:lt_vs_pt_teaser}
\end{figure}

In our view-independent case, we find that differentiable light tracing often outperforms path tracing in terms of optimization convergence behaviour (Fig.~\ref{fg:lt_vs_pt_teaser}).
Instead of transferring one radiance sample per path from a light source to a sensor, we update our radiance data at each ray-surface intersection and accordingly collect objective function derivatives from each of these locations in our adjoint tracing step, thereby using samples more efficiently.
To our knowledge, we present the first view-independent, analytically differentiable light tracing, inverse rendering method.
For comparisons to related work, we have implemented a reference differentiable path tracer working on our view-independent data structure.
We also compare to existing image-based methods in a baseline test scenario, by restricting our objective function to consider only surfaces that are visible by their camera.
In this way we separate the performance of the differentiation method from the benefits of operating on a larger global target.

One problem that has received a lot of attention recently is how to differentiate pixel values in the presence of discontinuous integrands due to silhouette edges or hard shadows.
Various strategies to resolve these issues have been proposed, such as reparametrizing the discontinuous integrands \cite{Loubet2019}, repositioning the samples around discontinuous edges \cite{Bangaru2020,Zeltner2021}, or separating the affected integrals into interior and boundary terms and applying separate Monte-Carlo estimators to each part \cite{Zhang2020,Yan2022}.
The main focus of these methods is to compute how a pixel value changes as a sharp edge or hard shadow moves \emph{within} this pixel.
Our light-tracing gradient formulation keeps ray-surface intersection points constant while moving light sources. Therefore, rays cannot move across feature edges when their source is perturbed, although lights could still move behind an occluding object.
The resulting moving shadows are currently not explicitly handled in our formulation; we leave them for future work.
Nevertheless, we achieve good optimization convergence using our gradients. While previous work often deals with situations where optimization parameters only affect a very small number of pixels, in our case, lighting parameters affect \emph{many} receiving 3d surfaces simultaneously, thereby diminishing the relevance of shadows moving through a few surface elements, see also \S\ref{sc:discontinuity}.
Our comparisons show that Mitsuba~3's reparametrization method, and PSDR's edge sampling, resolve discontinuities but introduce noise, leading to reduced optimization convergence rates compared to their standard version.
Using our spatio-directional data structure could in principle allow for extension of their code bases to the view-independent case, potentially including advanced discontinuity handling.
For the specific case of lighting design optimization, however, our adjoint formulation is more efficient than these methods (even before considering the overhead of discontinuity handling), as shown in Fig.~\ref{fg:imageCompBunnyOpt}-\ref{fg:mitsuba-cmp}.

\subsection{Lighting design}

The problem of designing lighting configurations has also been previously explored in computer graphics and tackled from different perspectives.
Many approaches use variations of the sketching metaphor, such as sketching the shape of shadows to indirectly move light sources~\cite{Poulin1997}, or painting highlights and dark spots to control light parameters in the scene~\cite{Shesh2007, Pellacini2007}.
Our view-independent differentiable rendering framework brings the sketching metaphor to 3d, providing a more intuitive way to paint the desired illumination directly into the scene.
Therefore, the user-painted objective is guaranteed to be free of contradictions that might otherwise occur when editing images from different views corresponding to the same 3d location.

Procedural or hierarchical optimization approaches to luminaire placement have also been explored~\cite{Lin2013,Schwarz2014,Gkaravelis2016,Jin2019}
As we focus on continuous optimization, these methods could be used to estimate a starting configuration before further optimizing the parameters of the generated light sources.

More recent approaches include the user in the optimization process~\cite{Sorger2016, Walch2019} directly, instead of automatically operating on a predefined target. These methods interactively display the current illumination and provide information about where and how this configuration could be improved. The user is then free to choose which measures to take in order to bring the scene closer to the desired state.
In the future, our approach could be integrated into such a user-centred editing framework to improve the design interaction between the user and the optimization system.

\section{Problem statement}

We start with the well-known rendering equation \cite{Kajiya1986}, which models the \emph{exitant} radiance  $L({\mathbf{x}},{\omega _o})$ from a point $\mathbf{x}$ on a surface in the outward direction $\omega _o$ as
\[L({\mathbf{x}},{\omega _o}) = {L_e}({\mathbf{x}},{\omega _o}) + \int_{{\mathcal{H}^2}} {f({\mathbf{x}},{\omega _i},{\omega _o}){(\mbf{n}\cdot \omega _i)}\;L_i({\mathbf{x}},{\omega _i})\;d{\omega _i}}. \eqn \label{eq:radiance} \]
The bidirectional reflectance distribution function (BRDF), ${f}$, encodes the material properties, and the \emph{incident} radiance $L_i({\mathbf{x}},{\omega _i})$ at $\mathbf{x}$ is related to the  radiance exitant from another point $\mathbf{x}'$ by ${L_i}({\mathbf{x}},{\omega _i}) = L({\mathbf{x}'}, {-\omega _i})$, such that $\mathbf{x}'$ is the nearest ray-surface intersection when tracing a ray from $\mathbf{x}$ in direction $\omega _i$.
Here, we focus on reflective light transport for brevity, but in principle, transmissive surfaces could be included in mostly the same way.
Consequently, we restrict the directional integration to the hemisphere ${\mathcal{H}^2}$ instead of the entire (unit) sphere ${\mathcal{S}^2}$. We also do not consider any volumetric light-scattering effects.

We then minimize an objective function that measures the quality of a given lighting configuration.
More formally, for this inverse rendering application, we consider objective functions $O$ of the form
\[O (L) = \frac{1}{2} \int_\Omega  { \frac{1}{{2\pi }} \int_{{\mathcal{H}^2}} {\alpha ({\mathbf{x}},{\omega _o}){{\left( {L({\mathbf{x}},{\omega _o}) - {L^*}({\mathbf{x}},{\omega _o})} \right)}^2}\;d{\omega _o}} d{\mathbf{x}}}, \eqn \label{eq:objectiveContinuous}\]
where $L^*$ is our illumination target, $\alpha$ is a user-defined weighting function, and $L({\mathbf{x}},{\omega _o})$ must satisfy the rendering equation.
As the emitted radiance $L_e(\mbf{p})$, due to a given set of light sources, depends on optimization parameters $\mbf{p}$, we look for local minima $\mbf{p}^*$:
\[\mbf{p}^* = \arg \min_{\mathbf{p}} O(L({\mathbf{p}})), {\text{~s.t.~} L \text{~satisfies Eq.~\eqref{eq:radiance}.}} \eqn \label{eq:optimProblem} \]

\begin{figure}[t]
\includegraphics[width=\columnwidth]{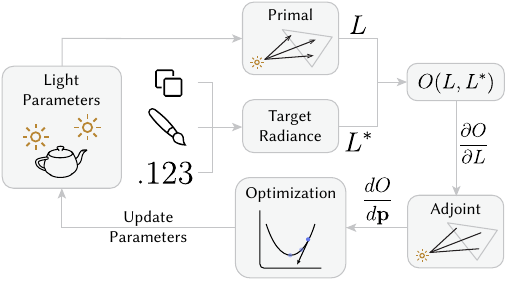}
\caption{
Illustration of our adjoint light-tracing optimization.
See also Fig.~\ref{fg:deriv-schematic} for further details on the primal and adjoint data flow.
}
\label{fg:overview-schematic}
\end{figure}

In order to address this problem, we compute an approximate solution, $L({\mathbf{x}},{\omega _o})$, on the surfaces of a virtual 3d scene in a \emph{primal} rendering pass (\S\ref{sc:primal}).
We then derive an \emph{adjoint} (or \emph{backpropagation}) rendering pass (\S\ref{sc:adjoint}) that allows us to compute derivatives of the optimization objective with respect to the parameters, $d O / d \mbf{p}$.
Consequently, we are able to apply gradient-based continuous optimization methods to the lighting design problem.

\section{View-independent adjoint light tracing} \label{sc:theory}

In the following sections, we first describe how we discretize the radiance data and optimization target (\S\ref{sc:discretization}), how we update this data while light tracing (\S\ref{sc:primal}), and finally how we compute the objective gradient through adjoint light tracing (\S\ref{sc:adjoint}).

\subsection{Spatio-directional data structure} \label{sc:discretization}
In order to represent the spatio-directional radiance field $L$ with a finite number of variables (or degrees of freedom, DOF), we first construct an appropriate interpolation scheme.
We then use this scheme to find approximate solutions to Eq.~\eqref{eq:radiance}.

For the spatial component, we choose a piece-wise linear interpolation, as often used in finite element (and specifically radiosity) methods \cite{Greenberg1986,Larson2013}:
\[L({\mathbf{x}},{\omega _o}) \approx \sum\nolimits_k {{L_k}({\omega _o}){\varphi _k}({\mathbf{x}})}. \eqn \label{eq:pwLin} \]
Instead of single nodal values, however, we consider $L_k$ to be a directional \emph{function} associated with the $k$-th mesh vertex and its nodal basis function $\varphi _k$.

For the directional component, we discretize each per-vertex function $L_k(\omega_o)$ using a hemi-spherical harmonic (HSH) basis \cite{Sillion1991,Gautron2004,Krivanek2005,Green2003}, specifically, the $2\pi$-normalized hemi-spherical harmonics ${H_m^l}$ \cite{Wieczorek2018}. Each directional function is consequently represented as
\[{L_k}({\omega _o}) \approx \sum\nolimits_{l = 0}^{l = n} {\sum\nolimits_{m =  - l}^{m = l} {{L_{klm}}{H_m^l}({\omega _o})} }, \eqn \label{eq:sh} \]
where $n$ is the maximal order used in this approximation, which implies that for each vertex $k$, we require $(n+1)^2$ directional coefficients.
Consequently, our spatio-directional approximation (using three colour channels, RGB) has $3 m (n+1)^2$ degrees of freedom, where $m$ is the number of mesh vertices.

For completeness, we summarize the construction of ${H_m^l}$ in the supplementary material. 
Furthermore, for the special case $n=0$, which is sufficient for diffuse surfaces, we effectively ignore the directional component and simply store vertex colours.
Note that even if we choose $n=0$, our light tracing method may still include glossy materials when simulating indirect lighting.

\begin{figure}[t]
\includegraphics[width=\columnwidth]{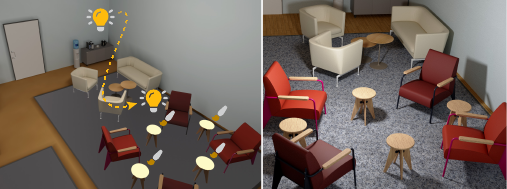}
\caption{ Iterative refinement of the scene shown in Fig.~\ref{fg:large-office}. First, desired changes are sketched on the 3d geometry (left), the optimization then adjusts the nearest light's position accordingly (right).
}
\label{fg:large-office-sketch}
\end{figure}

Our primary goal is to solve inverse problems based on this discretization, whereas we produce high-quality output images via standard rendering methods.
Therefore, we favour the simplicity of this discretization over alternatives such as texture-like bilinear interpolation or meshless bases \cite{Lehtinen2008}.
In contrast to previous work, we store exitant rather than incident radiance.
We also define our optimization objective on the exitant radiance, which can be intuitively painted by the user (Fig.~\ref{fg:large-office-sketch}).
Discretizing the target radiance field $L^*(\mbf{x},\omega_o)$ using the projection derived in the supplement, yields target coefficients ${L_{klm}^*}$ and subsequently a discrete analog of the objective function, Eq.~\eqref{eq:objectiveContinuous}:
\[O(L) =  \frac{1}{2} {\sum\nolimits_{klm} {{A_k}{\alpha_k}\alpha_l({L_{klm}} - L_{klm}^*) ^2}}, \eqn \label{eq:objective} \]
where $A_k$ is the area associated with vertex $k$, i.e,~one third of the sum of adjacent triangle areas. 
Here, we use a shorthand notation for the nested summation over vertices and HSH coefficients according to Eq.~\eqref{eq:pwLin} and \eqref{eq:sh}.
We choose to split the weights into a spatial and a directional component, $\alpha_k$ and $\alpha_l$ respectively.
Thus, the weights $\alpha_l$ adjust the influence of each HSH band: over-weighting lower bands for example would emphasize low-frequency components of the reflected radiance.
Similarly, $\alpha_k$ allows us to focus the attention of the optimizer on specific parts of the scene geometry.
We use a painting interface to allow the user to specify $L_{klm}^*$ as well as $\alpha_k$.
The HSH coefficients can be efficiently painted using a “directional” brush combined with an angular blur kernel, which is easy to compute because a closed-form solution to Laplacian smoothing exists in HSH space. 

For differentiable rendering, note that the derivative of Eq.~\eqref{eq:objective}, ${\partial O / \partial L_{klm}}$ is straightforward to evaluate, which we use in \S\ref{sc:adjoint}.

\subsection{Primal light tracing} \label{sc:primal}

\looseness=-1
One key feature of our discretization is the orthogonality of all basis functions.
Therefore, we can project each radiance sample computed during light tracing into our data structure quickly and independently.
In this section, we discuss how we solve the rendering equation using light tracing and update the coefficients $\{L_{klm}\}$ of our data structure accordingly.
We summarize this procedure in Alg.~\ref{alg:fwd}, and provide a  detailed derivation in the supplementary document.

\rev{Theoretically, we could regard our data structure as a generalized sensor and construct a path-space formulation of its measurement integral. Here, we instead consider light transport in the simpler particle tracing form, directly discretizing each light source's radiative power into a set of light rays. In \S\ref{sc:params} we then use ideas from the path-space theory to formulate the required derivatives for our light tracing approach.}{}
For every light source, we sample $N$ exitant rays, each representing a radiant flux $\Phi_r$ leaving the light source, distributed according to the light's emission profile, such that the sum of all these exitant samples equals the total radiative power of the light.
Let ${{\mathbf{x}}_0}$ be the origin of a ray. 
We then construct a light path up to a maximal number of indirect bounces, denoted as a sequence of ray-surface intersection points ${{\mathbf{x}}_1},\ldots,{{\mathbf{x}}_i},\ldots, {{\mathbf{x}}_\text{max}}$.

\begin{algorithm}[t]
\caption{Primal light tracing} \label{alg:fwd}
\begin{algorithmic}
\STATE Initialize all $L_{klm} = 0$
\FOR{ each light source with parameters $\mbf{p}$ }
 \FOR{ ray $r \in [1,N]$ }
 \STATE Generate exitant ray $(\mbf{x}_o,\; \omega_o)$ carrying flux $\Phi_r(\mbf{p})$
  \FOR{ $i \in [1,\text{maxBounces}]$ }
   \STATE Trace ray, find intersection point $(\mbf{x}_o,\; \omega_o) \rightarrow \mbf{x}_i$
   \FOR{ $N'$ local samples }
    \STATE $\omega_o \leftarrow$ uniform random in $\mathcal{H}^2$ 
    \FOR{ each vertex $k$ in the triangle containing $\mbf{x}_i$ }
     \FOR{ $l \in [0,n], \; m \in [-l,l]$ }
      \STATE Update coefficient $L_{klm}$ according to Eq.~\eqref{eq:directionalCoeffs}
     \ENDFOR
    \ENDFOR
   \ENDFOR
  \STATE Exitant ray:
  \STATE $\mbf{x}_o \leftarrow \mbf{x}_i$
  \STATE $\omega_o \leftarrow$ $\cos$-weighted random in $\mathcal{H}^2$ 
  \STATE $\Phi_r \leftarrow \Phi_r f(\mbf{x}_i,\omega_i,\omega_o) (\omega_o \cdot \mbf{n}_i) / {p_o}$
  \ENDFOR
 \ENDFOR
\ENDFOR
\end{algorithmic}
\end{algorithm}

The incident flux $\Phi_i$ arriving at each hit point ${{\mathbf{x}}_i}$ is  ${\Phi _1} = {\Phi _r}$ (direct illumination) and 
\[{\Phi _{i + 1}} = {\Phi _i}\;f_i(- {\omega _i},{\omega _{i + 1}})\;({\omega _{i + 1}} \cdot {{\mathbf{n}}_i})/{p_{i+1}}  \eqn \label{eq:indirectFlux}\]
(indirect illumination), where $f_i$ denotes the BRDF evaluated at ${{\mathbf{x}}_i}$, ${{\mathbf{n}}_i}$  is the surface normal at ${{\mathbf{x}}_i}$  and ${p_{i+1}}$  is the probability density of the exitant sample in direction ${\omega _{i +1}}  = ({{\mathbf{x}}_{i+1}} - {{\mathbf{x}}_{i}}) / \left\| {{{\mathbf{x}}_{i+1}} - {{\mathbf{x}}_{i}}} \right\|$.

At \emph{every} intersection point $\mbf{x}_i$, we sample the exitant radiance distribution due to the incident flux ${\Phi _i}$ along ${N'}$ randomly selected outward directions ${\omega _o}$.
We then update the coefficients $\{L_{klm}\}$, by adding the contribution of each sample ${\omega _o}$ to each vertex $k$, adjacent to $\mbf{x}_i$, as follows:
\[{L_{klm}} \leftarrow {L_{klm}} + \frac{1}{{N'}}\frac{1}{{{A_k}}} \; H_m^l({\omega _o}) {{\varphi _k}({\mathbf{x}_i})  \; f_i({-\omega _i},{\omega _o}) \; {\Phi _i}}, \eqn \label{eq:directionalCoeffs} \]
where $k$ refers to the vertices of the triangle containing ${{\mathbf{x}}_i}$, with vertex-associated area $A_k$, while $l$ and $m$ refer to each HSH basis function up to the selected maximal order $n$.
Please refer to our supplement for a detailed derivation of this update rule.

\begin{figure}[b]
\includegraphics[width=\columnwidth]{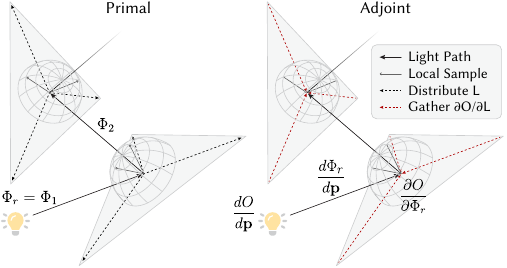}
\caption{Illustration of our primal and adjoint light tracing approach. In the primal pass, we trace radiant flux into the scene, while in the adjoint pass we collect derivatives of the objective function along the light path.
}
\label{fg:deriv-schematic}
\end{figure}

\subsection{Adjoint light tracing} \label{sc:adjoint}

So far we have discussed how we store the exitant radiance field (\S\ref{sc:discretization}) and how each light path affects the degrees of freedom of this field (\S\ref{sc:primal}).
We now turn our attention to improving the lighting configuration of a scene via gradient-based optimization and an adjoint formulation for computing the required gradients.

In contrast to previous work on image-based differentiable rendering, we do not handle derivatives along discontinuous edges in our method. This means we primarily ignore direct shadows, and instead focus on an efficient derivative formulation for the continuous case.
Our results show that these discontinuity problems have limited influence in common lighting design tasks; even in specifically designed test cases (Fig.~\ref{fg:simpleOffice_disc_exp} \& \ref{fg:cube_disc_exp}), the optimization algorithm sometimes recovers from receiving these biased gradients.

In the context of our light tracer, parameters that affect the lighting configuration may cause changes in brightness or colour ($\Phi_r$), as well as the ray origin, $\mbf{x}_0$.
In case the origin shifts (infinitesimally), the question is how this change propagates through the scene.
Does the first hit, $\mbf{x}_1$, also shift, and if so, what about the rest of the light path, $\mbf{x}_i$?
Automatic differentiation of Alg.~\ref{alg:fwd} would answer ``yes'' and compute derivatives of all hit points, ${ d \mbf{x}_i / d \mbf{x}_0 }$, as well as derivatives of the interpolation basis functions.
Our test cases (Fig.~\ref{fg:fd-naive-magnitude} \& \ref{fg:naiveGradComp}) show that this na\"ive approach does not lead to useful results.
Zhang et al.~\shortcite{Zhang2020} (PSDR) work around this issue by moving the evaluation to path space before applying AD.
Following a similar idea, we compute derivatives by treating the light path $\{\mbf{x}_i\}$ as a parameter-independent constant sample in path space when computing derivatives.
Consequently, if a light source moves, only the origin, $\mbf{x}_0$, and hence the first incident direction, $\omega_1 = {(\mbf{x}_1-\mbf{x}_0) / \| \mbf{x}_1-\mbf{x}_0 \| }$, are parameter dependent.
All other intersection points, $\mbf{x}_i$, remain fixed to the scene geometry. We first derive the general structure of our adjoint tracing method in this section, and then compute the required parametric derivatives in \S\ref{sc:params}.

Our goal is to compute the objective function gradient with respect to the parameters $\mathbf{p}$, which affect the solution $L_{klm}(\mathbf{p})$:
\[{({\nabla _{\mathbf{p}}}O )^\user1{T}} = \frac{{dO }}{{d{\mathbf{p}}}} = \sum\nolimits_{klm} {\frac{{\partial O }}{{\partial {L_{klm}}}}\frac{{d{L_{klm}}}}{{d{\mathbf{p}}}}}. \eqn \label{eq:objGrad} \]
The partial derivative ${\partial O / \partial {L_{klm}}}$ can be interpreted as the desired illumination change in the 3d scene per degree of freedom of the discretized radiance field.

We first consider the direct differentiation for $d {L_{klm}} / d \mbf{p}$ and then re-arrange the resulting terms to formulate an adjoint state per light path.
Every light path resulting from a primary light ray $r$ affects some coefficients of the solution ${L_{klm}}$ according to Eq.~\eqref{eq:directionalCoeffs}.
Consequently, we can split the sensitivity term $d {L_{klm}} / d \mbf{p}$ into contributions per path as follows:
\[ \begin{aligned}
  \left(\frac{{d{L_{klm}}}}{{d{\mathbf{p}}}}\right)_r & = \frac{{\partial {L_{klm}}}}{{\partial {\Phi _i}}}\frac{{\partial {\Phi _i}}}{{\partial {\Phi _r}}}\frac{{d{\Phi _r}}}{{d{\mathbf{p}}}}\quad &{\text{(flux)}} \\ 
   &+ \frac{{\partial {L_{klm}}}}{{\partial {\Phi _i}}}\frac{{\partial {\Phi _i}}}{{\partial f_1}}\frac{{\partial f_1}}{{\partial {\omega _1}}}\frac{{d{\omega _1}}}{{d{\mathbf{p}}}}\quad &{\text{(1\textsuperscript{st} bounce BRDF)}} \\ 
   &+ \frac{{\partial {L_{klm}}}}{{\partial f_i}}\frac{{\partial f_i}}{{\partial {\omega _1}}}\frac{{d{\omega _1}}}{{d{\mathbf{p}}}}\quad &{\text{(local BRDF)}} \\ 
\end{aligned} \eqn \label{eq:dLdp} \]
Note that for direct illumination, the direct incident flux $\Phi_1$ does not depend on the BRDF $f_1$, therefore the second term vanishes when $i=1$, as $\partial {\Phi_1}/\partial {f_1} = 0$.
Similarly, only $f_1$ depends on the direct incident angle ${\omega _1}$, but later BRDF evaluations do not, therefore, the third term vanishes for indirect illumination, because $\partial f_i/\partial {\omega _1} = 0$ for $i > 1$.
While we have omitted function arguments for brevity, it is important to point out that the \emph{direct} BRDF is sampled along local outward directions $\omega_o$, see Eq.~\eqref{eq:directionalCoeffs}, whereas the \emph{indirect} path evaluates the BRDF towards the exitant direction of the first bounce, $\omega_2$, Eq.~\eqref{eq:indirectFlux}.
Both of these terms depend on the first incident direction $\omega_1$, which changes when the origin of the light path moves.
This dependence is stated explicitly in Eq.~\eqref{eq:dLdp}; in the following we use the notation ${{{df_1({{\omega _2}})}}/{{d{\mathbf{p}}}}}$ and ${{{df_1({{\omega _o}})}}/{{d{\mathbf{p}}}}}$ to distinguish these terms.

\begin{algorithm}[t]
\caption{Adjoint light tracing} \label{alg:adjoint}
\begin{algorithmic}
\STATE Initialize $dO/d\mbf{p} = 0$
\FOR{ each light source with parameters $\mbf{p}$ }
 \FOR{ ray $r \in [1,N]$ }
 \STATE Generate exitant ray $(\mbf{x}_o,\; \omega_o)$
 \STATE Set $\partial O / \partial \Phi_r = 0$ and $\partial O / \partial f_1(\omega_2) = 0$
  \FOR{ $i \in [1,maxBounces]$ }
   \STATE Trace ray, find intersection point $(\mbf{x}_o,\; \omega_o) \rightarrow \mbf{x}_i$
   \IF{ $i == 1$ }
   \STATE Compute $d\Phi_r / d\mbf{p}$ (see \S\ref{sc:params})
   \STATE and $d f_1(\omega_2)/dp$
   \ENDIF
   \FOR{ $N'$ local samples }
    \STATE $\omega_o \leftarrow$ uniform random in $\mathcal{H}^2$ 
    \FOR{ each vertex $k$ in the triangle containing $\mbf{x}_i$ }
     \FOR{ $l \in [0,n], \; m \in [-l,l]$ }
      \STATE Update adjoint states according to Eq.~\eqref{eq:adjointFlux}
      \STATE and Eq.~\eqref{eq:adjointBRDF}
     \ENDFOR
    \ENDFOR
   \ENDFOR
  \STATE Exitant ray:
  \STATE $\mbf{x}_o \leftarrow \mbf{x}_i$
  \STATE $\omega_o \leftarrow$ $\cos$-weighted random in $\mathcal{H}^2$ 
  \ENDFOR
  \STATE Finally, update gradient according to Eq.~\eqref{eq:objGradAdjoint}.
 \ENDFOR
\ENDFOR
\end{algorithmic}
\end{algorithm}

Storing the full sensitivity matrix $d {L_{klm}} / d \mbf{p} = \sum\nolimits _r (d {L_{klm}} / d\mbf{p})_r$ by summation over contributions from all paths, Eq.~\eqref{eq:dLdp}, would consume a lot of memory as the number of parameters grows.
Our formulation avoids this memory cost and instead uses adjoint states \emph{per path}, which fit into the local memory of each GPU thread.

Instead of tracing derivatives of \emph{radiant flux} forward along a light path, we collect partial \emph{objective} derivatives \emph{backwards} along that path and compute a weighted sum of partial derivatives ${\partial O}/{\partial {\Phi _r}}$.
Nimier-David et al.~\shortcite{Nimier-David2020} also refer to a similar quantity as adjoint radiance in their work.
Their concept of adjoint radiance projects partial derivatives of the objective function from an image into the scene along camera paths, whereas we collect objective derivatives along light paths.
Combining Eq.~\eqref{eq:objGrad} and \eqref{eq:dLdp}, including summation over light paths noted above, and then re-arranging terms, yields:
\[\frac{{dO}}{{d{\mathbf{p}}}} = \sum\nolimits_r {\left( {\frac{{\partial O}}{{\partial {\Phi _r}}}\frac{{d{\Phi _r}}}{{d{\mathbf{p}}}} + \frac{{\partial O}}{{\partial f_1({{\omega _o}})}}\frac{{df_1({{\omega _o}})}}{{d{\mathbf{p}}}} + \frac{{\partial O}}{{\partial f_1({{\omega _2}})}}\frac{{df_1({{\omega _2}})}}{{d{\mathbf{p}}}}} \right)}. \eqn \label{eq:objGradAdjoint} \]
Here, ${\partial O}/{\partial {\Phi _r}}$ and ${\partial O}/{\partial f_1({{\omega _2}})}$, are \emph{adjoint states}, collecting objective function derivatives along the light path.
We update these terms at \emph{every} ray-surface intersection point $\mbf{x}_i$, analogous to the primal simulation, Eq.~\eqref{eq:directionalCoeffs}:
\[\frac{{\partial O}}{{\partial {\Phi _r}}} \leftarrow \frac{{\partial O}}{{\partial {\Phi _r}}} + \frac{{\partial O}}{{\partial {L_{klm}}}}\frac{{\partial {L_{klm}}}}{{\partial {\Phi _i}}}\frac{{\partial {\Phi _i}}}{{\partial {\Phi _r}}}, \eqn \label{eq:adjointFlux}\]
and
\[\frac{{\partial O}}{{\partial {f_1}({\omega _2})}} \leftarrow \frac{{\partial O}}{{\partial {f_1}({\omega _2})}} + \frac{{\partial O}}{{\partial {L_{klm}}}}\frac{{\partial {L_{klm}}}}{{\partial {\Phi _i}}}\frac{{\partial {\Phi _i}}}{{\partial {f_1}({\omega _2})}}. \eqn \label{eq:adjointBRDF}\]
Figure~\ref{fg:deriv-schematic} illustrates this idea, which we also outline in Alg.~\ref{alg:adjoint}.

Furthermore, the derivative wrt.~the BRDF under direct illumination, ${{{\partial O}}/{{\partial f_1({{\omega _o}})}}}$, only occurs at $\mbf{x}_1$ and can be  calculated directly from Eq.~\eqref{eq:directionalCoeffs} and \eqref{eq:objective}.
Similarly, the derivative of the BRDF itself wrt.~parameters $\mbf{p}$ follows directly from the BRDF's definition. 
Additionally, we find the derivative terms required to evaluate Eq.~\eqref{eq:adjointFlux} and \eqref{eq:adjointBRDF} as follows:
${\partial O / \partial L_{klm}}$ is the derivative of Eq.~\eqref{eq:objective}, while ${\partial L_{klm} / \partial \Phi_i}$ follows from differentiating Eq.~\eqref{eq:directionalCoeffs}.
Finally, ${\partial \Phi_i / \partial \Phi_r}$ and ${\partial \Phi_i / \partial f_1(\omega_2)}$ are tracked along the light path, effectively expanding the recursive product induced by Eq.~\eqref{eq:indirectFlux} and collecting factors.

Note that in our formulation, primal and adjoint tracing are entirely separate rendering passes.
In contrast, automatic differentiation correlates primal rendering and derivative calculation, and previous methods work around this issue by running the primal pass twice (``\emph{decorrelation}'').
Here, whether we evaluate the adjoint step in a \emph{correlated} or \emph{decorrelated} way is simply a question of whether we choose the same or a different random seed compared to the primal step.
In validation tests and finite difference comparisons, we use the correlated evaluation (i.e.,~keeping the same random seed in all rendering passes), while for all other results we use decorrelated evaluation.

\subsection{Optimization parameters} \label{sc:params}

The final step to evaluating the objective function gradient according to Eq.~\eqref{eq:objGradAdjoint} is computing $d\Phi_r/d\mbf{p}$ for all parameters $\mbf{p}$ of the lighting configuration.
Note that this is the only term that depends on the type of each light source (point, spot, area, etc.) and on the meaning of each parameter (position, intensity, rotation, etc.).
As $\Phi_r$ denotes the direct illumination, these derivatives can generally be calculated analytically, and we briefly summarize them here, while deferring details to the supplementary document.

\subsubsection{Intensity and colour}
Intensity (per colour channel) ${I_e}$ for point-shaped lights (or analogously emissive power for area lights) directly affects the emitted flux per light ray, hence the corresponding derivative is straightforward to calculate.
We find it useful to parametrize intensity with a quadratic function, $I_e = 0.5 \mbf{p}_j^2$, where $\mbf{p}_j$ is the corresponding optimization parameter, to prevent negative values.
If the task is to optimize chromaticity, but maintain a constant intensity, we need to find a colour vector $\mbf{c} = {(r, g, b)}$ such that $r + g + b = 1$.
We implement this constraint by projecting $\mbf{c}$ to $\mbf{c}_p = \mbf{c} / (r + g + b)$ in each optimization step, and similarly projecting the derivative by multiplying with the Jacobian $(d{\mathbf{c}_p}/d{{\mathbf{c}}})$.

\begin{figure}[t]
\includegraphics[width=\columnwidth]{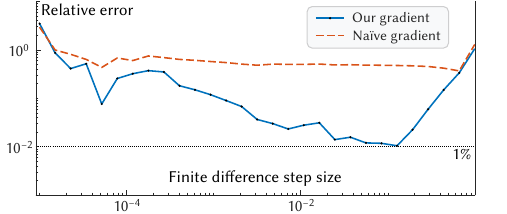}
\caption{
Validation of our gradient formulation using finite difference approximations at various step sizes (for the light's 3d position in Fig.~\ref{fg:simpleOffice-opt-merge}a).
Our gradient closely matches FD approximations at adequate step sizes and we observe the expected trade-off between floating-point errors (too small FD steps) and approximation errors (too large FD steps).  The na\"ive version, however, exhibits large systematic errors across a wide range of step sizes.
}
\label{fg:fd-naive-magnitude}
\end{figure}

\subsubsection{Position and rotation}
\label{sc:geometricDerivs}

Let us now consider the derivative of the objective function with respect to a light's position $d O / d \mbf{x}$. As discussed in \S\ref{sc:adjoint}, the na\"ive approach would compute the set of spatial derivatives $d\mbf{x}_i / d\mbf{x}$ for the entire path. Automatic differentiation also produces this result, see Mitsuba~3's `ptracer' in Fig.~\ref{fg:imageCompBunnyOpt} and \ref{fg:imageCompBunnyGradVis}.

\begin{figure}[t]
\includegraphics[width=\columnwidth]{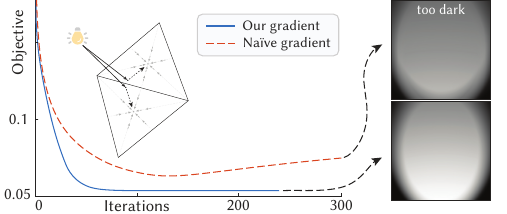}
\caption{Optimizing the position of a spot light such as to find a trade-off between uniformity and brightness on a Lambertian plane.
The na\"ive gradient calculation (red dashes) \emph{fails} to place the light at the correct distance from the plane.
The inset illustrates how interpolation basis function gradients in the na\"ive approach push each ray towards the centroid of the intersected triangle, in order to increase brightness equally at all corners.
Our method (blue line), in contrast, provides improved gradient information, enabling the optimizer to find the best distance. }
\label{fg:naiveGradComp}
\end{figure}

There is, however, a serious issue with this na\"ive approach:
computing derivatives in this way generally \emph{fails} to produce reliable descent directions that would be useful for gradient-based optimization of lighting configurations, Fig.~\ref{fg:naiveGradComp}. This issue arises because any spatial derivative of $\mbf{x}_i$ results in a vector in the plane of the triangle containing $\mbf{x}_i$, with discontinuous jumps between triangles, as illustrated in the inset of Fig.~\ref{fg:naiveGradComp}.
Note that this type of discontinuity problem is distinct from the case of moving hard shadows due to occlusions (we intentionally ignore the latter).
The former issue, however, causes severe limitations for differentiable light (or particle) tracing and must be addressed.

The PSDR approach \cite{Zhang2020} avoids this issue by evaluating the \emph{entire} path according to the path-integral formulation (including all geometric terms explicitly) and then applying automatic differentiation.
In contrast to their general formulation, we deal specifically with parameters that affect light sources, which means that we only need to consider the first geometric term, while our adjoint states handle the indirect light path.
For the remaining direct part, we now apply a similar idea in explicitly differentiating the geometric term at $\mbf{x}_1$.
For improved performance, we compute these derivatives analytically\rev{,}{ (with the help of symbolic math tools),} rather than relying on \emph{code-level} AD.
The resulting expressions fit into thread-local GPU memory allowing for efficient parallel computation.

Our solution to finding useful gradients wrt.~\emph{position and rotation} of a light source is to consider the ``reverse'' direction of the primary light ray ($\mbf{x}_0 \rightarrow \mbf{x}_1$)\rev{}{.
Effectively, we compute the derivatives of the direct illumination represented by $\Phi_r$ of an arriving ray,} \emph{as if} we were tracing the ray in the other direction (via next event estimation from the scene geometry to the light source).
Consequently, we treat $\mbf{x}_1$ as constant, while the light source moves (or rotates) and consider a ``virtual'' intensity per ray, ${{\tilde I_e} = {\Phi_r r^2 / (\omega_1 \cdot \mbf{n})}}$, representing the irradiance this ray transports from the light source to $\mbf{x}_1$.
The direct illumination due to $\Phi_r$ behaves like ${\tilde I_e (\omega_1 \cdot \mbf{n}) /  r^2}$ locally.
Consequently, the derivative of the incident flux with respect to the light's position becomes
\[ \frac{d \Phi_r }{ d \mbf{x}} ={ \tilde I_e \; \frac{ d((\omega_1 \cdot \mbf{n})/{r^2}) }{ d \mbf{x}}}. \eqn \label{eq:posDeriv}\]
We compute this derivative using symbolic math software, as shown in our supplementary document.

\looseness=-1
Figure~\ref{fg:fd-naive-magnitude} compares errors of our gradient to the na\"ive version discussed earlier, relative to finite difference (FD) approximations of various step sizes.
As expected, for small step sizes finite differencing suffers from floating-point errors, whereas for large step sizes, FD approximation error dominates.
For a suitable range of step sizes, our gradients agree well with finite difference approximations.
The na\"ive approach, on the other hand, shows systematic error over a wide range of FD step sizes.
This error accumulates during the summation over many rays; for single light paths, we have verified the correct implementation of the na\"ive gradient calculation.
In Fig.~\ref{fg:naiveGradComp} we compare our improved approach to the na\"ive version on the basic optimization task of positioning a spot light. 
Using our method, gradient descent optimization converges quickly, and reliably remains in the optimal configuration.
We intentionally choose a standard gradient descent approach (i.e.,~updating ${\mathbf{x}} \leftarrow {\mathbf{x}} - \alpha (dO/d{\mathbf{x}})$) instead of more elaborate optimization algorithms for this example, because it most clearly exposes the flaws of the na\"ive differentiation approach.

Moving on to rotations, we parametrize the orientation of light sources by a rotation vector ${\mbf{\theta }}$, which defines a rotation matrix ${\mathbf{R}(\mbf{\theta })}$ following Rodrigues' formula.
The world-space orientation (normal $\mathbf{n}$ and tangent $\mathbf{t}$) results from applying this rotation to the material-space normal and tangent $(\mbf{n}_0, \mbf{t}_0)$ respectively.
We use the linear small-angle approximation to avoid numerical issues when $\|\mbf{\theta }\|$ approaches zero.
Applying the chain rule, we find the derivative of the flux wrt.~the rotation vector:
\[\frac{{d{\Phi _r}}}{{d{\mathbf{\theta }}}} = \frac{{\partial {\Phi _r}}}{{\partial {\mathbf{n}}}}\frac{{d{\mathbf{n}}}}{{d{\mathbf{\theta }}}} + \frac{{\partial {\Phi _r}}}{{\partial {\mathbf{t}}}}\frac{{d{\mathbf{t}}}}{{d{\mathbf{\theta }}}}, \eqn \label{eq:rotVecDeriv}\]
where we compute the terms $d\mbf{n}/d\mbf{\theta}$ and $d\mbf{t}/d\mbf{\theta}$ using symbolic differentiation of Rodrigues' formula.
Finally, we find ${{\partial \Phi_r }/{\partial {\mathbf{n}}}}$ and ${{\partial \Phi_r }/{\partial {\mathbf{t}}}}$ via symbolic differentiation, again keeping $\mbf{x}_1$ fixed.
The exact expressions depend on the type of light source.
For instance, IES data files \cite{IESNA2020}---which are widely used in architecture---define the emitted intensity by tabulated values, interpolated bilinearly over the unit sphere; spot lights with soft edges use a quadratic attenuation profile between the inner and outer cone angle; whereas area light sources assume a cosine-weighted emission profile.
Please refer to our supplementary document for further details.
Note that for \emph{area lights} the ray origin moves when the light is rotated, i.e.,~${\mbf{x}_0(\mbf{\theta})}$.
(For our purposes we only consider rigid rotation and translation.)
In this case, the derivatives wrt.~the light's orientation include an additional term, analogous to Eq.~\eqref{eq:posDeriv} to account for the shift of the ray origin, as described in the supplement.

\begin{figure}[t]
\includegraphics[width=\columnwidth]{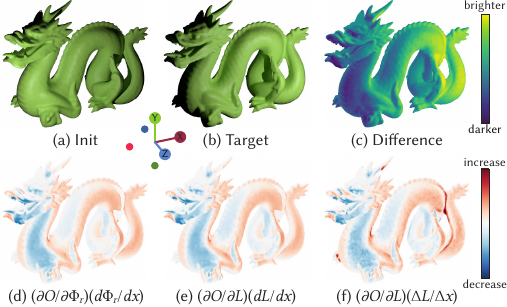}
\caption{
Visualizing gradients on the Stanford Dragon: the top row shows the initial (a) and intended (b) lighting, as well as their difference (c).
The second row shows gradient contributions for the $x$ coordinate of a point light using our adjoint approach (d), direct differentiation (e), and a finite difference approximation (f). Image subtitles indicate the calculation; BRDF derivatives are taken into account but not explicitly stated here for brevity.
}
\label{fg:grad-vis-dragon}
\end{figure}

\subsubsection{Further parameters}
Because Eq.~\eqref{eq:objGradAdjoint} separates derivatives of the objective function from derivatives of the lighting configuration, we can in principle extend our optimization system to include other parameters relatively easily.
The main question we have to address for each parameter, is how it affects the incident radiance (discretized using many rays) at the first ray-surface intersection, thereby formulating derivatives of each ray's radiant flux, such as to represent the expected change in the direct illumination.
In general, we believe this idea of including knowledge about a global behaviour into local, but discretized, derivative calculations could also be beneficial for other applications.

\section{Visualizing gradient contributions} \label{sc:gradVis}

In common camera-based differentiable rendering, the gradient of the rendered image with respect to a specific parameter can be visualized as a \emph{gradient image} or a \emph{gradient texture} \cite{Zhang2020,Nimier-David2020,Zeltner2021} when optimizing an object's appearance. Our objective function instead evaluates the full radiance field instead of a single image.
During optimization, we employ an adjoint formulation that \emph{avoids} computing derivatives of this radiance field explicitly.
We can compute these derivatives for visualization and comparison, but more interestingly, we can also use our per-ray adjoint states for a different kind of visualization showing \emph{adjoint gradient contributions} as follows.

For any point $\mbf{q}$ on the surface geometry (that is directly visible from a selected light source), we construct a light path that originates at that light and passes through $\mbf{q}$ as its first ray-surface intersection, and then continues bouncing through the scene.
Evaluating Eq.~\eqref{eq:objGradAdjoint} restricted to these paths, we find contributions to the derivative $dO/d\mbf{p}_j$ that ``flow'' through the point $\mbf{q}$.
Intuitively, we can think of each point $\mbf{q}$ casting a vote on how the parameter $\mbf{p}_j$ should change in order to improve the objective function (while taking indirect reflections from $\mbf{q}$ into account).
Note that this quantity is different from \emph{adjoint radiance} \cite{Nimier-David2020}, which does not include derivatives wrt.~parameters, or indirect bounces.

\looseness=-1
The example in Fig.~\ref{fg:grad-vis-dragon} shows which parts of the dragon would improve the objective function if the light moves right (red) or left (blue).
The final objective function gradient $d O / d \mbf{p}$ can be thought of as the sum of these contributions.
We compare our visualization to a reference implementation that directly differentiates the radiance field and a finite difference approximation in Fig.~\ref{fg:grad-vis-dragon}, and also to related work in Fig.~\ref{fg:imageCompBunnyGradVis}.
For (mostly) direct lighting, our method and the reference yield nearly identical results;
when the objective is strongly influenced by \emph{indirect} illumination, however, our adjoint visualization draws gradient contributions at the first ray-surface intersection, whereas direct differentiation produces a response at the target surface instead (see the  additional example in our supplement).

\begin{figure}[t]
\includegraphics[width=\columnwidth]{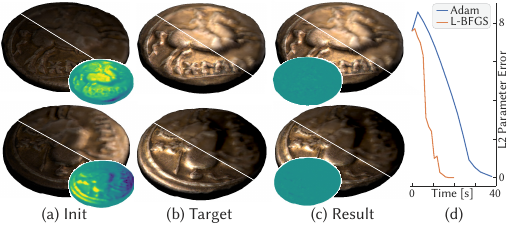}
\caption{
Recovering the ground-truth position of a point light illuminating a glossy coin. initial lighting (a), the target (b), and our result (c); both  ADAM and L-BFGS reliably converge to the correct result (d).
The insets show differences to the target. Top row: front view, second row: side view.
The top-right parts in (a, b, c) show our data structure, while the bottom-left shows a high-quality rendering of the same lighting.}
\label{fg:coin-ground-truth}
\end{figure}

\begin{figure}[t]
\includegraphics[width=\columnwidth]{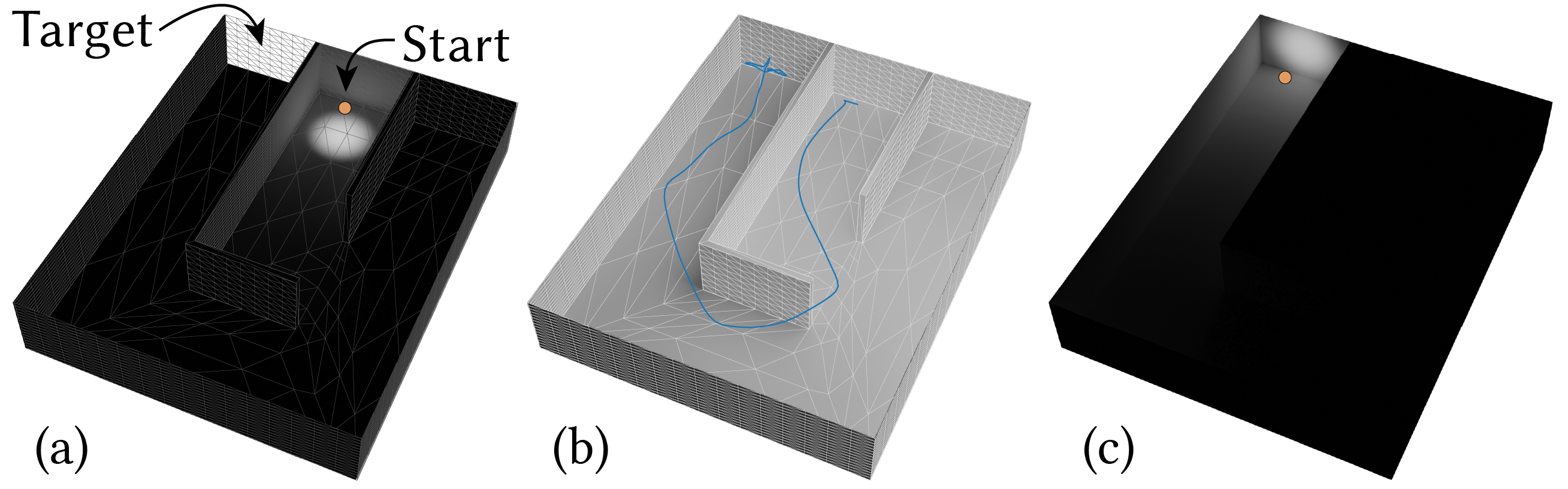}
\caption{
\looseness=-1
Optimizing indirect illumination: a spot light placed in a simple labyrinth (a) should illuminate the target (overlaid on the far left wall), which is only reached by indirect light. The ADAM optimizer simultaneously rotates and moves the light source around the corners (b) to find the ``exit'' (c).}
\label{fg:indirectOpt}
\end{figure}

\section{Results}

In this section we first cover verification test cases, as well as comparisons to related methods; we then demonstrate the ability of our approach to address creative lighting design cases.
Unless stated otherwise, we trace two indirect bounces in all of our results.
Our light transport simulations run on the GPU, specifically a \emph{NVIDIA GeForce RTX 3080}, using hardware-accelerated ray tracing via the Vulkan API.
We also evaluate the objective function on the GPU, but transfer the optimization parameters and their derivatives to the CPU, in order to access publicly available optimization libraries.
Running the optimization algorithm on the CPU is not a performance bottleneck, as the size of the parameter vector is much smaller than the size of the radiance data.

In general, predicting \emph{which} optimization algorithm will perform best for a specific task is difficult. In our results we demonstrate that we provide gradient information that works with \emph{multiple} commonly used optimization methods, most notably gradient descent, ADAM, and L-BFGS \cite{Nocedal1980}.
Our implementation uses the LBFGS++ library~\shortcite{Qiu2021}, as well as the ADAM algorithm of Kingma and Ba~\shortcite{Kingma2014} (using decay rates $\beta_1 = 0.9$ and $\beta_2 = 0.999$ as suggested in the original publication for all our results).
For comparisons to gradient-free CMA-ES, we use the code by Hansen~\shortcite{Hansen2003,Hansen2021}.

\begin{figure}[t]
\includegraphics[width=\columnwidth]{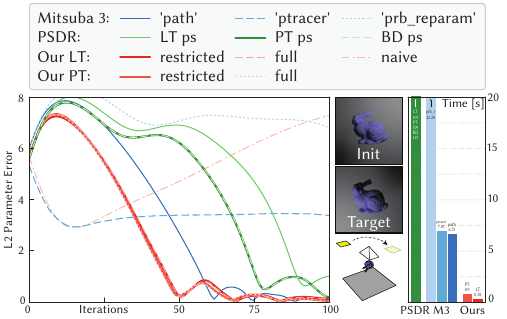}
\caption{Baseline test case: moving an area light to a ground-truth position (inset images), showing the convergence of our method, PSDR and Mitsuba~3 in terms of parameter-space distance. All methods run ADAM with step size $\alpha = 0.25$, $4$ indirect illumination bounces and $1$~million samples. 
On this simple scene, restricting our objective function to visible surfaces (`restricted') results in the same convergence rate as using the entire radiance data (`full').
Note that Mitsuba's `prb\_reparam' method eventually converges (around $150$~iterations), whereas their `ptracer' does not. The graph on the right shows the total runtime for $100$~iterations: our light tracer is ca.~twice as fast as our reference path tracer, while comparing the three Mitsuba integrators shows the runtime impact of discontinuity handling.
}
\label{fg:imageCompBunnyOpt}
\end{figure}

\subsection{Verification tests}

We first verify that our differentiable light-tracing system is capable of solving various inverse problems, where a well-defined solution exists. 
We perform ground-truth recovery tests, where the illumination target is copied from the result corresponding to a specific lighting configuration in various scenes: a large office (Fig.~\ref{fg:large-office}), a glossy coin, using hemi-spherical harmonics up to order $n=5$ (Fig.~\ref{fg:coin-ground-truth}), as well as the Stanford bunny (Fig.~\ref{fg:imageCompBunnyOpt}).
For these examples, we visualize the differences to the ground-truth target before and after optimization, or compare convergence of the optimization to related methods (see details in \S\ref{sc:mitsuba}).

Apart from ground-truth tests, we verify that our method correctly finds solutions for well-defined objectives on specific surfaces, Fig.~\ref{fg:naiveGradComp} and Fig.~\ref{fg:indirectOpt}, and compare our gradient calculation to finite difference approximations in Fig.~\ref{fg:fd-naive-magnitude}, Fig.~\ref{fg:grad-vis-dragon}, and Fig.~\ref{fg:simpleOffice-opt-merge}.
In our supplementary material we also provide a visual comparison of different orders of HSH interpolation, which shows that our radiance data structure converges under refinement.

Finally, we test the capabilities of our system in dealing with indirect illumination.
In Fig.~\ref{fg:indirectOpt}, we place a spot light into a small ``labyrinth'', such that light can only reach the ``exit'' via multiple bounces.
The side walls of the labyrinth have a dark target value to push the light away from the walls, with a very low weight ($\alpha_k = 0.004$), whereas the exit has a very bright target with unit weight to attract the light source.
We trace light paths with $10$ indirect bounces in this example.
Optimizing for position and rotation simultaneously using ADAM with step size $\alpha = 0.2$ successfully navigates the light through the labyrinth in ca.~$10$~s and $200$ gradient evaluations.
In principle, L-BFGS also successfully solves this problem, but the momentum estimated by ADAM results in a smoother, visually appealing motion of the light source.

\begin{figure}[t]
\includegraphics[width=\columnwidth]{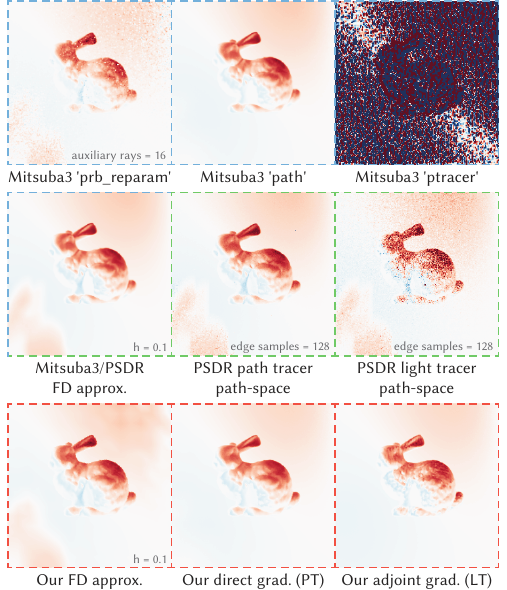}
\caption{Gradient images wrt.~the $x$ coordinate of the area light for the initial state of Fig.~\ref{fg:imageCompBunnyOpt} (primal rendering of this state and the target inset there);
colour range: $-0.05$ (blue) to $+0.05$ (red). 
Each image uses $16$~million primary samples, additional samples for discontinuity handling are stated per image as applicable.
Note that Mitsuba 3's particle tracer fails to produce correct results. PSDR's path-space approach works, but light tracing causes more noise than their path tracer. Our gradient calculations ignore the moving shadow (bottom left corner), but produce comparable results to Mitsuba's standard path tracer. In particular, our adjoint light tracing produces equally clean results, while other light tracers are noticeably noisier.
}
\label{fg:imageCompBunnyGradVis}
\end{figure}

\subsection{Comparison to previous work}
\label{sc:mitsuba}

Current work on inverse rendering always considers images to define the optimization objective.
We establish a baseline for comparing our (camera-free) view-independent approach to existing methods by \emph{restricting} our objective function to data available to image-based methods (camera-visible surfaces) on a test scene, where one camera sufficiently captures most surfaces (except of course the back of the bunny, Fig.~\ref{fg:imageCompBunnyGradVis}).
In this test, we use ideally diffuse materials for all objects in the scene, and no directional data ($n=0$) for our light tracer and reference path tracer, to allow for a fair comparison\rev{}{, i.e.,~the directional radiance field contains no additional information over the camera view}. 
As shown in Fig.~\ref{fg:imageCompBunnyOpt}, our adjoint light tracing matches state-of-the-art methods in this setting in terms of optimization convergence, while our adjoint formulation, combined with an efficient GPU-based implementation outperforms other methods in terms of runtime. We use an equal number of samples for these comparisons. Finally, note that for the unrestricted case (`full') with $n=0$, Fig.~\ref{fg:imageCompBunnyOpt} also contains an ``object-space'' baseline comparison between our light tracer and reference path tracer, which shows that light tracing ($0.38~s$) outperforms path tracing ($0.9~s$) for an equal number of samples (delivering equal optimization convergence).

Mitsuba's light tracing method (`ptracer'), using automatic differentiation, fails to converge, similar to our na\"ive differentiation approach in Fig.~\ref{fg:naiveGradComp}.
The differentiable path-integral formulation of PSDR\footnote{Most features of the PSDR code are currently only available in their CPU-based implementation. PSDR currently only supports area lights, but neither point, spot, nor IES lights.} \cite{Zhang2020}, on the other hand, produces good results in this simple setting for both light and path tracing (with slightly more noise and reduced convergence rate of their light tracer).

\begin{figure}[t]
\includegraphics[width=\columnwidth]{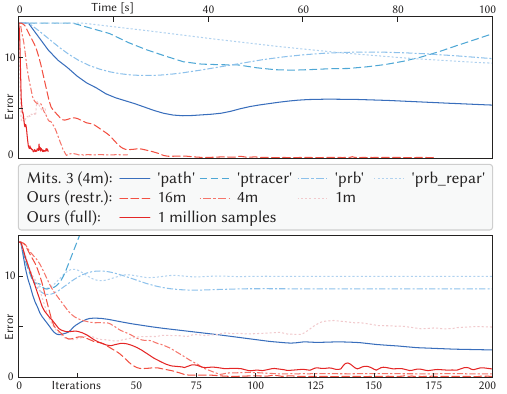}
\\ \vspace{2mm}
\includegraphics[width=0.95 \columnwidth]{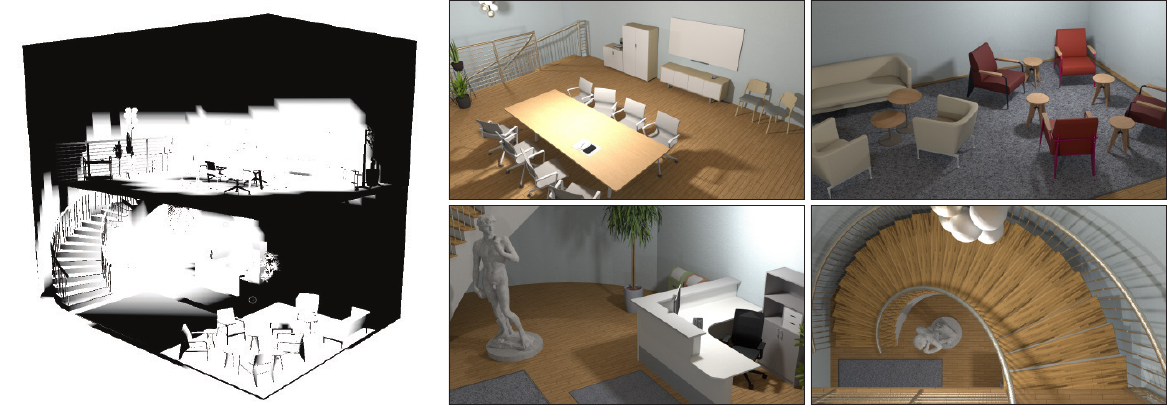}
\caption{
Runtime (top) and convergence (middle) comparisons of our method with Mitsuba~3 on the same scene as shown in Fig. \ref{fg:large-office}, modified with four cameras and four lights sources. We show results for a restricted version of our method consisting only of camera-visible data (bottom, left), as well as a full version. Four different integrators of Mitsuba 3 were used and all Mitsuba 3 tests were done with $4$ million samples, while for our algorithm we chose 1, 4, and 16 million samples. The error is the Euclidean distance between current and target light positions and intensities in parameter space. Note that our full version with just 1 million samples performs similar to our restricted version using 4 times the amount of samples.
Bottom row: we restrict our objective by weighting the contribution of vertices (left, white: $\alpha_k = 1$, black: $\alpha_k = 0$) based on their visibility to the cameras used by Mitsuba. The camera views used when optimizing with Mitsuba are shown on the right.
}
\label{fg:mitsuba-cmp}
\end{figure}

While related image-based methods (Mitsuba, PSDR) handle discontinous edges in various ways, they also introduce additional noise into the gradient (compared to Mitsuba's standard AD path tracer), causing reduced optimization convergence (Fig.~\ref{fg:imageCompBunnyOpt}). As discontinuity handling also comes with a runtime cost, we choose speed over (theoretical) accuracy for the scope of this paper; we analyze the effects of the resulting gradient bias in \S\ref{sc:discontinuity}. Overall, our approach produces less noisy gradient data compared to existing differentiable light tracing methods.

\begin{figure}[t]
\includegraphics[width=\columnwidth]{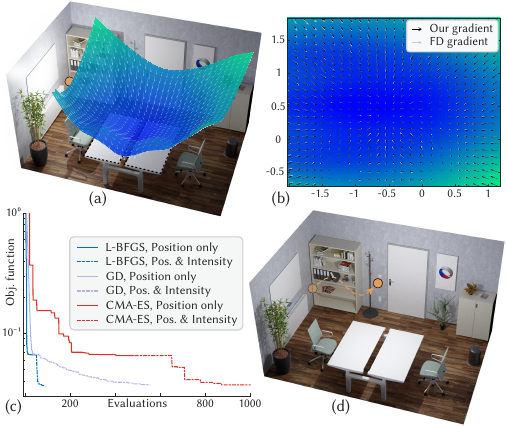}
\caption{ \looseness=-1
A basic lighting design case:
the target requires the tables (dashed lines in a) to be lit uniformly (the rest of the scene does not affect the objective $\alpha_k=0$).
We plot the objective function on a regular grid in the horizontal plane (a), where the orange dot marks the initial position of the light source, producing the rendered result in the background.
Arrows show negative gradients wrt.~the in-plane position of the light, which reliably point towards a local minimum. We visually compare these gradients to a finite difference approximation (b).
Initial and optimal light position overlaid on the final illumination (d): the light becomes brighter and moves centrally over the tables during optimization.
We compare convergence of gradient descent and L-BFGS (both using our gradients) to gradient-free CMA-ES optimization (c).
With each method, we first optimize only the light's 3d position (solid lines) and then its position and intensity combined (dashed lines).}
\label{fg:simpleOffice-opt-merge}
\end{figure}

\begin{figure}[t]
\includegraphics[width=\columnwidth]{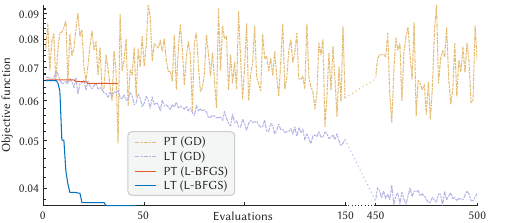}
\caption{
Equal-time comparison between our adjoint light tracing and our reference differentiable path tracing implementation on the combined optimization of the light source's position and intensity, as in Fig.~\ref{fg:simpleOffice-opt-merge}.
Using our gradients, both gradient descent (step size $\alpha = 0.1$) and L-BFGS converge robustly, whereas differentiable path tracing produces more noisy gradients, preventing convergence. (Note that we squeeze the x-axis between iterations $150-450$ to show the later gradient descent iterations. L-BFGS, on the other hand, terminates when it fails to find an improved configuration.)
}
\label{fg:simpleOfficeLTvsPT}
\end{figure}

For further comparison, we also implement a reference differentiable path tracer following the ideas of Stam~\shortcite{Stam2020} on top of our view-independent data structure: we uniformly sample each triangle and trace paths with next event estimation to find the incident radiance.
Finally, we sample exitant directions locally, apply the BRDF and update our data structure according to Eq.~\eqref{eq:directionalCoeffs}.
This path tracer uses the same sampling strategies to construct indirect illumination paths as our light tracer for a fair comparison of our light tracing method against ``object-space'' path tracing.
The gradients produced by this path tracer closely match Mitsuba's AD path tracer (Fig.~\ref{fg:imageCompBunnyGradVis}).
For the baseline test scene we observe similar convergence behaviour between restricted objective function data and full view-independent data (i.e.,~camera-visible vs.~all surfaces in the scene), as intended, while our light-tracing approach is substantially faster than path tracing.
We show an extension of this baseline test in the supplement, where adding an off-camera box around the scene highlights the advantage of our view-independent approach.
\rev{}{Our reference path tracer also achieves convergence similar to our light tracer on the intensity optimization test (Fig.~\ref{fg:ground-truth}).} 
Furthermore, Fig.~\ref{fg:simpleOfficeLTvsPT} shows an equal-time comparison on the more complex optimization problem in Fig.~\ref{fg:simpleOffice-opt-merge}, where our light-tracing method yields improved convergence rates, due to reduced (gradient) noise (similar behaviour is also visible in Fig.~\ref{fg:lt_vs_pt_teaser}, as well as the indirect illumination ``labyrinth'' example in our video).

\begin{figure*}[t]
\includegraphics[width=\textwidth]{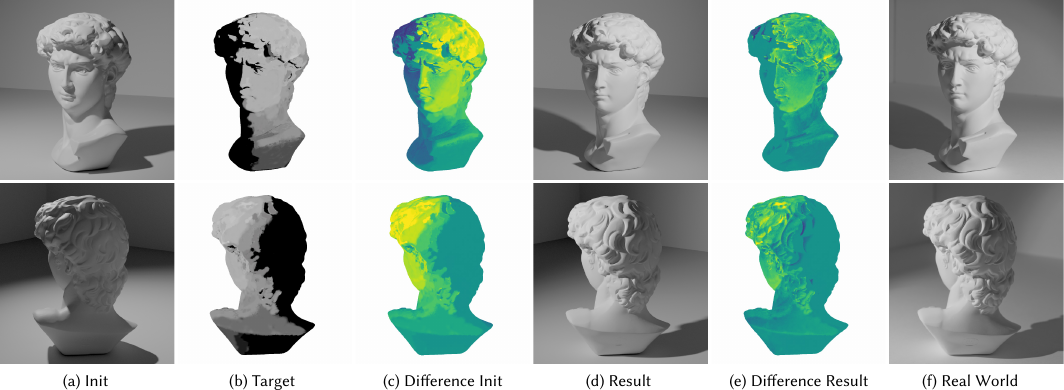}
\caption{ Lighting design study on the bust of David: we first generate a flat initial lighting of the laser-scanned model (a), then sketch a desired side lighting in 3d (b) and optimize the position and rotation of the point light source to match this target (d). We then re-create the optimized lighting setup using the real-world specimen (f). Columns (c) and (e) show the difference to the target before and after optimization respectively.
}
\label{fg:real-world-david-bust}
\end{figure*}

Finally, we compare our approach to Mitsuba~3 on a large office scene (Fig.~\ref{fg:mitsuba-cmp}), where Mitsuba uses four cameras to cover most of the scene.
We again restrict our method to surfaces visible by any camera for comparison (Fig.~\ref{fg:mitsuba-cmp}).
Mitsuba's reparmetrization shows reduced optimization performance (likely due to increased noise in the gradient estimates).
Using our method without restriction to visible surface improves convergence, even at lower sample count.
This behaviour is due to the global coupling between parameters in the lighting design problem, which is in this sense a more challenging optimization problem than, for instance, optimizing texture colours given a target image.

In summary, in terms of optimization convergence, our adjoint light tracing method at least matches, and on complex scenes outperforms, state-of-the art inverse rendering systems on the lighting design problem.
Our formulation enables a fast GPU implementation, which results in far better runtime performance.

\begin{figure}[t]
\includegraphics[width=\columnwidth]{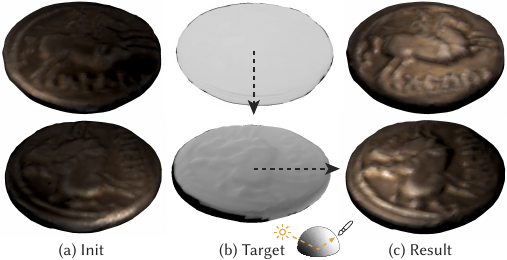}
\caption{
Lighting an ancient coin according to a user-defined directional target (top row: front view, bottom row: side view).
Initially, the light source is positioned on the left side of the coin (a) when viewed from the front.
The user-defined target (b) aims for lighting mostly towards the camera in the front view. Our result (c) matches the desired lighting direction.
}
\label{fg:coin-manual-target}
\end{figure}

\subsection{Lighting design applications}

\subsubsection{Small office lighting}

We first show an example of a fully automatic optimization of a single point light in a small office scene, Fig.~\ref{fg:simpleOffice-opt-merge}.
The optimization target specifies bright, but uniform lighting on the top surface of both tables, a common regulatory requirement for work spaces.
In the initial configuration, a point light is placed off to one side, causing the work space to be too dimly and unevenly lit.
We test different optimization methods on two subsequent tasks in this scene: the first optimizes the 3d position of the light, placing it centrally above the tables for a good trade-off between brightness and uniformity; the second jointly optimizes for the light's intensity and position.
The optimal light placement is now just underneath the ceiling (thereby improving uniformity), with the intensity increased to compensate for the larger distance (Fig.~\ref{fg:simpleOffice-opt-merge}d).

We compare the performance of GD and L-BFGS to gradient-free CMA-ES \cite{Hansen2003} in Fig.~\ref{fg:simpleOffice-opt-merge}c.
In the first, position-only, optimization task, both L-BFGS and GD reliably find a good solution in $6.5~s$ and $42$ objective and gradient evaluations, or $7.9~s$ and $52$ evaluations respectively.
In comparison, CMA-ES requires around $200$ objective evaluations to find a good solution (initial standard deviation $\sigma = 1$ for all parameters).
In general, the computation time required to evaluate gradients via an adjoint method is close to the time needed to compute the solution itself.
Therefore, even though CMA-ES does not need to evaluate gradients, the total runtime is still significantly slower than gradient-based methods.
See also Table~\ref{tb:comparison} for timings of the primal and adjoint evaluations in our examples.
In the second task, when we optimize for position and intensity simultaneously, the Hessian approximation built by the L-BFGS method captures the relation between the height (distance) and intensity.
Consequently, L-BFGS finds a good solution (Fig.~\ref{fg:simpleOffice-opt-merge}) in $8.2~s$ and $45$ evaluations.
Plain gradient descent and ADAM (as shown in our video), on the other hand, converge more slowly, while CMA finds an acceptable solution after about $400$ evaluations.

\begin{figure*}[t]
\includegraphics[width=\textwidth]{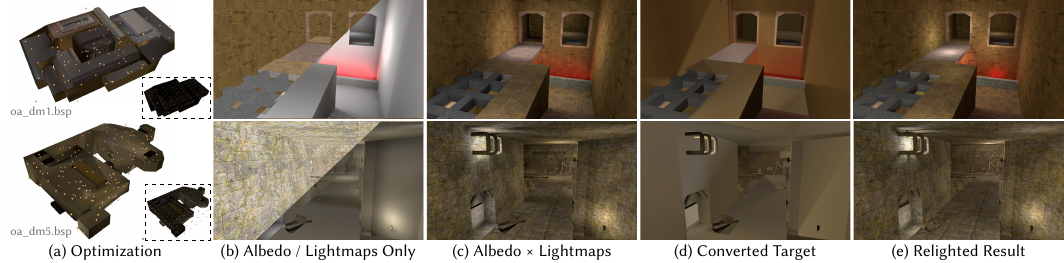}
\caption{ Recovering light parameters for relighting an old PC game, OpenArena~\shortcite{OpenArena2008}, which originally shipped with baked lighting only: the original ``look'' of the game (c) is produced by combining albedo textures and lightmaps (b). From this data, we construct an optimization target (d) and use it to find a new lighting configuration (e) as close as possible to the original (c). Our optimization starts from a regular grid layout of lights and finds a suitable arrangement for them (a). Ultimately, this enables relighting old games often missing dedicated lights with more advanced rendering techniques (e.g., ray-tracing).
}
\label{fg:q3details}
\end{figure*}

\subsubsection{Directional lighting of a glossy coin}

In Fig.~\ref{fg:coin-manual-target} we show an example where directional lighting plays an important role.
The user specifies a direction-dependent target by painting directly into the HSH-discretized data structure and coefficients ${L_{klm}^*}$.
In this example, we set the per-channel weights $\alpha_l$ such as to under-weight the undirected lighting component represented by $H_0^0$ by a factor of $0.1$ relative to the directional components of the radiance data.
Using our adjoint gradients, an L-BFGS optimizer successfully navigates the light source around the coin to find the best matching directional lighting configuration.

\subsubsection{Real-world bust of David}

We demonstrate the applicability of our system to an artistic lighting design process on a small bust of David figurine, Fig.~\ref{fg:real-world-david-bust}.
We first laser-scan the bust and build a 3d mesh suitable for our simulation using a Metris MCA~3600 articulated arm with a Metris MMD~50 laser scanner.
In our interactive user interface, we then sketch a desired shading onto this model.
Performing $100$ iterations of ADAM optimization ($\alpha = 0.14$), which takes about $7.5$~s, we find the position and orientation of a spot light that closely matches the painted target.
In order to validate our result, we replicate this lighting configuration on the real-life specimen, Fig.~\ref{fg:real-world-david-bust}f.
We use a Cree\textregistered{} XLamp\textregistered{} CXA1304 LED and a custom-made snout to limit the cone angle to $45°$, matching our simulation.
We apply only basic white balance and gradation curve adjustments to the real-world photographs.

\subsubsection{Refurbishing baked lighting}

\looseness=-1
Another interesting application of our system is ``refurbishing'' old video games.
In many cases, the diffuse (potentially also indirect) illumination is baked into static \emph{lightmaps}.
However, all information about the original light sources, which would be required to render the scenes with a modern real-time ray tracing method, is often lost.
Here, we reconstruct light sources such that the given textures (assumed to represent albedo), combined with the recovered lighting produce a similar impression as the original game.
We demonstrate this approach in Fig.~\ref{fg:q3details}, where we build lighting configurations for two different scenes \rev{from OpenArena~\shortcite{OpenArena2008}.}{}
In each case, we first initialize a regular grid \rev{(e.g., $9 \times 9 \times 2$)}{} of low-intensity point lights, covering the bounding box of the scene.
We then optimize for position, colour, and intensity of all lights simultaneously \rev{(up to $972$ parameters)}{} using ADAM.
Figure~\ref{fg:q3details} shows overviews of the initial and final lighting configuration, and interior views of our results.

\begin{figure}[t]
\includegraphics[width=0.95\columnwidth]{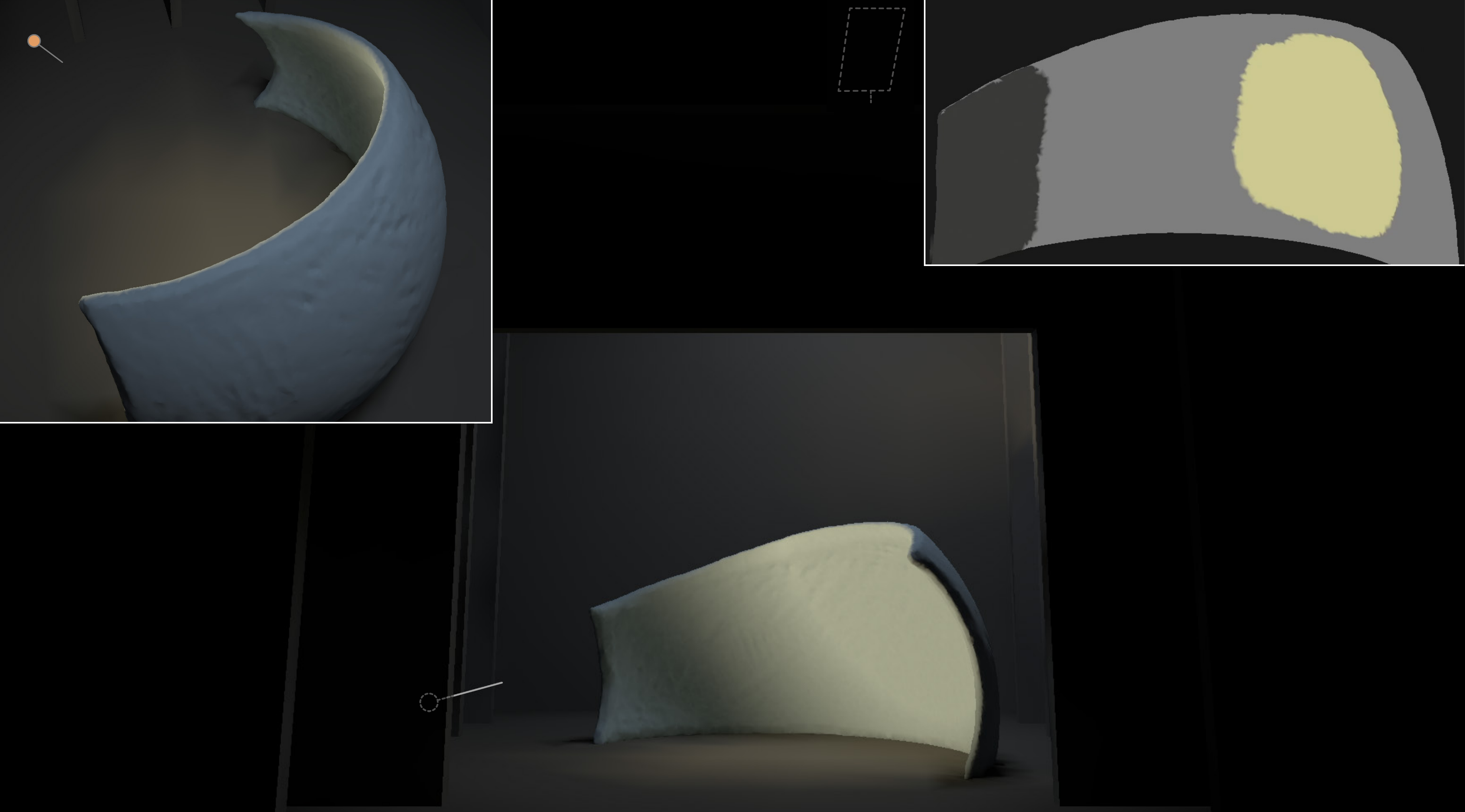}
\caption{
Lighting a theatre stage; please also refer to our accompanying video for the design interaction.
The main image shows the audience view, illuminated by a spot light on the left and an area light on top.
The inset on the left shows a top-down view of the stage.
The inset on the right shows a part of the user-drawn lighting target.
}
\label{fg:theatre}
\end{figure}

\subsubsection{Large office}

In a more complex example, we show a lighting optimization of a large office consisting of two floors.
This scene is illuminated by two point lights (one over the staircase, one over the lounge area), two spot lights near the reception and statue, an area light over the conference table, and two highly anisotropic lights (using IES intensity profiles \cite{IESNA2020}) on the top floor.
We first show a result that recovers a given ground-truth (Fig.~\ref{fg:large-office}) and compare the optimization convergence to our reference path tracer (Fig.~\ref{fg:lt_vs_pt_teaser}).
We also compare our method to Mitsuba~3 on this scene (replacing the IES lights, which are not supported by Mitsuba, with spot lights) in \S\ref{sc:mitsuba}.
We then demonstrate the ability of our system to interactively edit the desired illumination and automatically reconfigure the lighting design accordingly in Fig.~\ref{fg:large-office-sketch}.

\subsubsection{Theatre stage lighting}

Finally, we show an interactive design session, where the user starts from a completely unlit scene and iteratively adds light sources, selects optimization parameters, and updates the illumination target.
The interactive workflow alternates between user manipulations and automatic lighting optimization to realize the intended design, as shown in our video.
Figure~\ref{fg:theatre} summarizes the result of this design example.
In total, this lighting design session required about $10$~minutes to complete, with just over $2$~minutes spent on automatic design optimization, and the remaining time on various user interactions including visual inspection of the results.
Please also refer to Table~\ref{tb:comparison} for a performance overview of our results.

\begin{figure}[t]
 \includegraphics[width=\columnwidth]{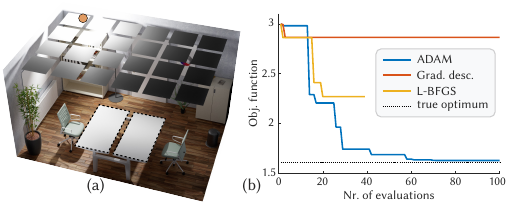}
 \caption{
 Scene from Fig.~\ref{fg:simpleOffice-opt-merge} with additional occluders: the light is placed above the ceiling (a) and must only move horizontally to best illuminate the tables (dashed lines), using the same objective as before.
 Gradient descent and L-BFGS fail to converge (b), while ADAM is more robust due to its momentum term.
 We find the true optimum by gradient-free CMA-ES optimization for comparison.
}
\label{fg:simpleOffice_disc_exp}
\end{figure}

\begin{figure}[t]
\includegraphics[width=\columnwidth]{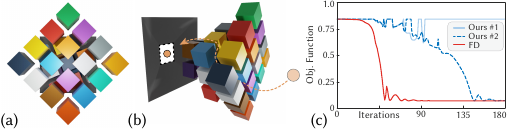}
\caption{
In this test scene, containing a staggered array of occluding cubes (a), the light needs to navigate through this arrangement of cubes to illuminate the uniformly white target on the back wall (b).
Depending on the random seed during rendering, ADAM optimization (step size $\alpha=0.05$) may get the light stuck inside of a cube, but occasionally manages to find a way through, due to momentum (``spikes'' in the objective function graph (c) indicate that the light is inside of a cube).
We also optimize (with the same settings) using the central finite difference approximation of gradients (FD-step $h = 0.03$) for reference, which results in a smooth navigation of the light in between the cubes.
}
\label{fg:cube_disc_exp}
\end{figure}

\subsection{Discontinuity Failure Cases} \label{sc:discontinuity}
In this section, we analyze specific situations where our method may fail to converge properly due to not handling occlusion discontinuities (moving shadows).
First, we show a modified lighting design example in Fig.~\ref{fg:simpleOffice_disc_exp}, where the optimizer should navigate the light (in the horizontal plane) in between the occluding panels on the ceiling of a small office.
In a second test case, we construct a scene exhibiting a high degree of occlusion, Fig.~\ref{fg:cube_disc_exp}, where the light must move through a staggered array of cubes.
In both cases, we clearly observe that our gradients are lacking information about how the moving occlusion boundaries (shadows) affect the objective function, causing most optimization attempts to fail.
Especially gradient descent or L-BFGS exhibit convergence problems in these tests, as they rely on being provided with a valid descent direction.
Adding momentum, as in ADAM, on the other hand, can sometimes recover and "skip over" areas exposed to problematic gradients.
In fortunate cases, where momentum compensates for gradient bias, ADAM sometimes finds acceptable solutions, even with inaccurate gradient data.

One additional problem caused by noisy or wrong gradients is the increased chance of a light entering a closed-off scene object, effectively trapping it and stalling optimization.
Note, however, that lights passing through a wall cause a truly non-differentiable discontinuity that could still occur even with known discontinuity handling methods and should be addressed by continuous collision detection instead.
We leave this line of investigation for future work.

\begin{table}[b]
\caption{Overview of our results. Columns: scene reference, optimization algorithm, step size $\alpha$, number of primary rays (threads) $N$, number of mesh vertices $m$, average time per primal and adjoint evaluation $t_{L}$, $t_{\nabla}$ [ms], total wall clock time for the entire optimization $t$ [s] (`*' marks average of multiple optimizations).
} \label{tb:comparison}
\begin{tabular}{llrrrr}
\toprule
Scene (Fig.) & Optim. & $\alpha$ & $N$; $m$ & $t_{L}$; $t_{\nabla}$ & $t$ \\
\midrule
Large Office (\ref{fg:large-office}, \ref{fg:lt_vs_pt_teaser}) & ADAM   & 0.1    & 14.7e6 & 127    & 64.0 \\
\ (light trace)       &        &        & 1 409 381 & 86    &  \\
Large Office (\ref{fg:lt_vs_pt_teaser}) & ADAM   & 0.1    & 39.7e6 & 107    & 66.4 \\
\ (path trace)       &        &        & 1 409 381 & 110    &  \\
Coin (\ref{fg:coin-ground-truth}) & L-BFGS & 1      & 4.2e6 & 147    & 18.4 \\
\ (HSH $n=5$) &        &        & 6 345  & 174    &  \\
Labyrinth (\ref{fg:indirectOpt}) & ADAM   & 0.2    & 1.0e6 & 31     & 10.8 \\
       &        &        & 1 893  & 22     &  \\
Labyrinth & ADAM   & 0.2    & 2.4e6 & 35     & 9.5 \\
\ (path trace, video) &        &        & 1 893  & 12     &  \\
Small Office (\ref{fg:simpleOffice-opt-merge}) & L-BFGS & 1      & 4.2e6 & 152    & 14.7 \\
       &        &        & 301 009 & 28     &  \\
Small Office (\ref{fg:simpleOfficeLTvsPT}) & L-BFGS & 1      & 75e6 & 127    & 13.5 \\
\ (path trace) &        &        & 301 009 & 62     &  \\
Small Office & L-BFGS & 1      & 4.2e6 & 673    & 16.2 \\
\ (HSH $n = 3$) &        &        & 301 009 & 61     &  \\
David (\ref{fg:real-world-david-bust}) & ADAM   & 0.14   & 4.2e6 & 28     & 6.8 \\
       &        &        & 535 071 & 37     &  \\
Coin painted (\ref{fg:coin-manual-target}) & L-BFGS & 1      & 4.2e6 & 123    & 22.5 \\
\ (HSH $n=5$) &        &        & 6 345  & 154    &  \\
OpenArena (\ref{fg:q3details}) & ADAM   & 0.15    & 40.5e6 & 654    &  72.5 \\
       &        &        & 8100 & 42    &  \\
Theatre stage (\ref{fg:theatre}) & L-BFGS & 1      & 18e6 & 97     & *13.5 \\
\ (5x L-B. + 1x A.) & ADAM   & 0.2    & 38 448 & 202    & 60.6 \\
\bottomrule
\end{tabular}

\end{table}

\section{Limitations and future work}

We focus on continuous optimization of a given lighting configuration, relying on the user to provide an initial placement of light sources, either interactively or as part of the original scene description.
In the future we will investigate extending our method with mixed-integer optimization approaches to also optimize for non-continuous parameters like the selection of light sources. 
Currently, we do not handle discontinuities due to moving shadows in our derivatives, which has been done in previous work on camera-based differentiable rendering.
Because our lighting objectives measure relatively large areas and our radiance solution is continuously interpolated, our method converges robustly nonetheless.
In the future, it will be interesting to investigate discontinuity handling for specific applications, such as lampshade design, that rely on accurately placing an occluding object in front of a light source.
In our implementation, we do not employ advanced sampling strategies, like importance sampling, which are often used to increase the efficiency of Monte Carlo methods.
Such methods have recently been successfully applied to differentiable image rendering.
As the resulting methods may require parameter-dependent sampling, thereby complicating the calculation of derivatives, we leave this line of research in the context of our light-tracing approach for future work.
Similarly, we currently use a user-defined number of HSH basis functions to represent the directional component of the radiance field.
In the future it could be interesting to investigate choosing the interpolation order adaptively based on the material properties and lighting conditions.

\section{Conclusion}

In summary, our method enables lighting optimization via differentiable rendering, for the first time providing two important features: interactive feedback and easily modifiable, view-independent objectives.
We present a novel, analytical adjoint light-tracing formulation, rather than relying on automatic differentiation.
Our view-independent radiance data structure can be quickly updated, both during light tracing and while painting illumination targets.
Combining our adjoint formulation and data structure yields an efficient implementation, which improves per-iteration runtime (equal samples) and convergence rates, compared to existing methods.
Providing objective function values and gradients in each iteration allows us to use \emph{any} first-order optimizer in a black-box fashion.
Note that we do not use interpolation-basis-function derivatives in our approach, therefore gradient accuracy does not suffer due to low quality meshes.

Our validation tests show that differentiating the incident flux, while keeping the first ray-surface intersection point constant, captures gradients of information that is contained in the distribution of discrete light rays.
Our method computes more accurate gradients compared to a na\"ive approach as evidenced by finite difference and optimization convergence tests.
We also show that (in equal time comparisons) adjoint light tracing results in improved optimization convergence behaviour than differentiable path tracing, and that our method outperforms existing image-based differentiable rendering methods on a baseline test scene.
We also provide a novel visualization of \emph{adjoint} gradient contributions to analyze the composition of the objective function gradient.
Furthermore, we show the applicability of our system to various lighting design tasks that cannot be easily handled by state-of-the-art image-based inverse rendering, including large-scale work spaces, artistic installations, and video game refurbishing.

\balance 
\begin{acks} 
We thank our students Mathias Schwengerer and Matthias \hbox{Preymann} for helping with our implementation, our colleagues  Balint Istvan Kovacs and Ingrid Erb for their help on the theatre stage example, as well as Henry Ehlers for narrating our video.\\

The Iberian Coin model, Fig.~\ref{fg:coin-ground-truth} and \ref{fg:coin-manual-target} has been created by `Itagues' on
\href{https://sketchfab.com/itagues}{Sketchfab}. The scenes shown in Fig.~\ref{fg:large-office} and \ref{fg:simpleOffice-opt-merge} were created with pCon.planner \cite{pcon2023} using assets from the pCon Catalog by
Vitra, Bene, SIGEL, Object Carpet, EFG, Abstracta, Frovi, about OFFICE, Cascando, Aarsland, Gotessons, Hailo and Bosig.
The model of Michelangelo's `David', Fig.~\ref{fg:large-office}, was created by `3DWP' on \href{https://sketchfab.com/3dwp}{Sketchfab}.

The results in Fig.~\ref{fg:q3details} use geometry, texture, and lightmap data from \href{https://openarena.ws}{OpenArena} maps by Tim Willits, Bob ``dmn\_clown'' Isaac, and
Conrad ``anyone'' Colwood, available under CC-BY-SA licence.

This research was funded by the Austrian Science Fund (FWF) project F 77 (SFB “Advanced Computational Design”).
\end{acks}

\bibliographystyle{ACM-Reference-Format}
\bibliography{references}


\ifsupplement

\clearpage
\section*{Supplementary material}
\title{Supplementary material}

\setcounter{page}{1}
\setcounter{equation}{0}
\renewcommand\theequation{S.\arabic{equation}}
\setcounter{figure}{0}
\renewcommand\thefigure{S\arabic{figure}}
\setcounter{table}{0}
\renewcommand\thetable{S\arabic{table}}
\setcounter{algorithm}{0}
\renewcommand\thealgorithm{S\arabic{algorithm}}

\subsection*{(Hemi-) Spherical Harmonics}
The spherical harmonics (eigenfunctions of the Laplacian on the unit sphere, see also \cite{Green2003,Wieczorek2018}) are defined as
\[Y_l^m(\varphi ,\theta ) = \left\{ {\begin{array}{*{20}{l}}
  {K_l^m P_l^{ - m}(\cos \theta )\sqrt 2 \sin ( - m\varphi ),}&{m < 0} \\ 
  {K_l^m P_l^m(\cos \theta ),}&{m = 0} \\ 
  {K_l^m P_l^m(\cos \theta )\sqrt 2 \cos (m\varphi ),}&{m > 0} 
\end{array}} \right. \eqn\]
where $\varphi$ is the azimuth angle, $\theta \in [0, \pi]$ is the elevation angle, and the scaling factors are
\[K_l^m = \sqrt {{{(2l + 1)}}\frac{{(l - \left| m \right|)!}}{{(l + \left| m \right|)!}}}. \eqn\]

The associated Legendre polynomials $P_l^m$ are commonly written as recurrence relations (Green~\shortcite{Green2003} also provides code for evaluating them):
\[\begin{gathered}
  (l - m)P_l^m(x) = x(2l - 1)P_{l - 1}^m(x) - (l + m - 1)P_{l - 2}^m(x), \hfill \\
  P_m^m(x) = {( - 1)^m}(2m - 1)!!{(1 - {x^2})^{m/2}}, \hfill \\
  P_{m + 1}^m(x) = x(2m + 1)P_m^m(x). \hfill \\ 
\end{gathered} \eqn \]

The hemispherical harmonics $H_l^m$ are eigenfunctions of the Laplacian on the unit hemisphere.
They are defined in almost the same way as $Y_l^m$ (see also \cite{Gautron2004}), except that the elevation angle is limited to the range $[0, \pi/2]$, and the associated Legendre polynomials are evaluated for the argument ${(2 \cos \theta -1)}$ instead of ${(\cos \theta)}$.
The local coordinate system $(\varphi, \theta)$ is oriented according to the surface normal and tangent.

\subsection*{Derivation of the radiance cache update rule}

In this section, we summarize the derivation of the update rule presented in the main paper as Eq.~\eqref{eq:directionalCoeffs}.
First, note the following properties of our spatial and directional interpolation basis functions $\varphi$ and $H_m^l$, namely nodal Kronecker and partition of unity spatially, and $2\pi$-orthonormality for the hemispherical harmonics:
\[{\varphi _k}({{\mathbf{x}}_j}) = {\delta _{jk}},\quad \sum\nolimits_k {{\varphi _k}({\mathbf{x}})}  = 1, \eqn \label{eq:appdxLinBasis} \]
\[\int_{{\mathcal{H}^2}} {H_m^l(\omega )H_{m'}^{l'}(\omega )\;d\omega  = 2\pi {\delta _{mm'}}{\delta _{ll'}},\quad H_0^0 = 1}. \eqn \label{eq:appdxHshProp} \]

Assume, for now, an a-priori known radiance field ${L(\mbf{x},\omega_o)}$, which we discretize according to Eq.~\eqref{eq:pwLin} and \eqref{eq:sh}, i.e.~we aim to find the coefficients $L_{klm}$ best approximating this field.
In the spatial domain, we determine the local directional functions ${L_k}({\omega _o})$ in Eq.~\eqref{eq:pwLin} via $L_2$-projection with mass lumping, in line with a finite element analysis \cite{Larson2013}.
$L_2$-projection of a function $f$ to its approximation $f_h$ is defined by 
$\int  {(f - {f_h}){\varphi _k}\;d{\mathbf{x}}}  = 0$.
Applying this definition to the radiance field and discrete approximation, Eq.~\eqref{eq:pwLin}, yields
$\int_\Omega  {L({\mathbf{x}},{\omega _o}){\varphi _k}({\mathbf{x}})\;d{\mathbf{x}}}  = \int_\Omega  {\sum\nolimits_j {{\varphi _j}({\mathbf{x}})} {L_j}({\omega _o}){\varphi _k}({\mathbf{x}})\;d{\mathbf{x}}} $.
Mass lumping then identifies $L_j$ with $L_k$, and by partition of unity the right-hand side simplifies to
${L_k}({\omega _o}) \int_\Omega  {{\varphi _k}({\mathbf{x}})d{\mathbf{x}}} = {{L_k}({\omega _o}) A_k}$.
Consequently, we find
\[{L_k}({\omega _o}) = \frac{1}{{{A_k}}}\int_{{\Omega _k}} {{\varphi _k}({\mathbf{x}})L({\mathbf{x}},{\omega _o})\;d{\mathbf{x}}}, \eqn \label{eq:spatialRep}\]
where we integrate over the one-ring neighbourhood ${\Omega _k}$ of vertex $k$ (i.e.~the support region of the shape function $\varphi _k$), and $A_k = |{\Omega _k}|/3$ is the vertex-associated area.
Intuitively, Eq.~\eqref{eq:spatialRep} computes the area-weighted average of the exitant radiance around each vertex.

\begin{figure}[t]
\def\svgwidth{\columnwidth}
\begingroup%
  \makeatletter%
  \providecommand\color[2][]{%
    \errmessage{(Inkscape) Color is used for the text in Inkscape, but the package 'color.sty' is not loaded}%
    \renewcommand\color[2][]{}%
  }%
  \providecommand\transparent[1]{%
    \errmessage{(Inkscape) Transparency is used (non-zero) for the text in Inkscape, but the package 'transparent.sty' is not loaded}%
    \renewcommand\transparent[1]{}%
  }%
  \providecommand\rotatebox[2]{#2}%
  \newcommand*\fsize{\dimexpr\f@size pt\relax}%
  \newcommand*\lineheight[1]{\fontsize{\fsize}{#1\fsize}\selectfont}%
  \ifx\svgwidth\undefined%
    \setlength{\unitlength}{1926.17806502bp}%
    \ifx\svgscale\undefined%
      \relax%
    \else%
      \setlength{\unitlength}{\unitlength * \real{\svgscale}}%
    \fi%
  \else%
    \setlength{\unitlength}{\svgwidth}%
  \fi%
  \global\let\svgwidth\undefined%
  \global\let\svgscale\undefined%
  \makeatother%
  \begin{picture}(1,2.22096142)%
    \lineheight{1}%
    \setlength\tabcolsep{0pt}%
    \put(0,0){\includegraphics[width=\unitlength,page=1]{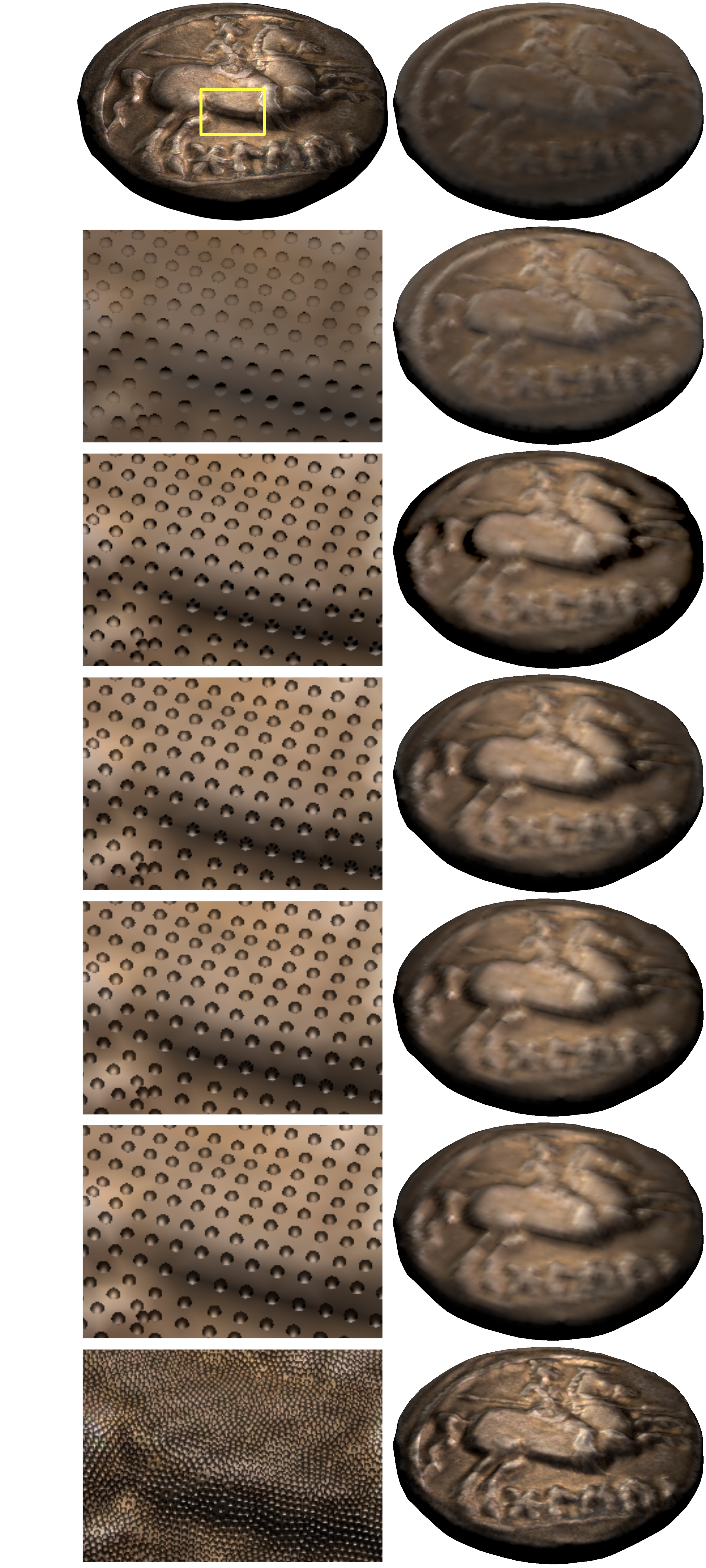}}%
    \put(-0.00237878,1.71777285){\makebox(0,0)[lt]{\lineheight{1.25}\smash{\begin{tabular}[t]{l}$n=1$\end{tabular}}}}%
    \put(-0.00471505,1.40058395){\makebox(0,0)[lt]{\lineheight{1.25}\smash{\begin{tabular}[t]{l}$n=3$\end{tabular}}}}%
    \put(-0.00504353,1.08386957){\makebox(0,0)[lt]{\lineheight{1.25}\smash{\begin{tabular}[t]{l}$n=5$\end{tabular}}}}%
    \put(-0.00562764,0.76602354){\makebox(0,0)[lt]{\lineheight{1.25}\smash{\begin{tabular}[t]{l}$n=7$\end{tabular}}}}%
    \put(-0.00540856,0.44872508){\makebox(0,0)[lt]{\lineheight{1.25}\smash{\begin{tabular}[t]{l}$n=9$\end{tabular}}}}%
    \put(-0.00540856,0.13233773){\makebox(0,0)[lt]{\lineheight{1.25}\smash{\begin{tabular}[t]{l}$n=9$ \\ $2$ subd.\end{tabular}}}}%
  \end{picture}%
\endgroup%

\caption{ 
Convergence of our spatio-directional radiance data structure with increasing order of the hemi-spherical harmonic basis: top row shows a high-quality rendering (left) and a result without any directional data (i.e.~$n=0$). The left column shows a close-up of the region indicated by the yellow rectangle with per-vertex HSH functions visualized on small hemispheres.
The last row shows the result after applying two LS3 subdivision steps to all triangles.
 }
\label{fg:hsh-coins-fwd}
\end{figure}

In the directional domain, we follow the same idea, applying $L_2$-projection over the hemisphere $H^2$:
$\int_{{H^2}} {{L_k}({\omega _o})H_{m'}^{l'}({\omega _o})\;d{\omega _o}}  = \sum\nolimits_{lm} {{L_{klm}}} \int_{{H^2}} {H_m^l({\omega _o})H_{m'}^{l'}({\omega _o})\;d{\omega _o}}$.
We then take advantage of the orthogonality of the hemi-spherical harmonic basis functions to determine the coefficients of our solution:
\[{L_{klm}} = \frac{1}{{2\pi }}\int_{{\mathcal{H}^2}} {H_m^l({\omega _o})\;{L_k}({\omega _o})\;d{\omega _o}}. \eqn \label{eq:directionalRep}\]

Consequently, the projection operator for a given radiance field onto our disretize data structure is
\[{L_{klm}} = \frac{1}{{{A_k}}}\frac{1}{{2\pi }}\int_{{\Omega _k}} {\int_{{H^2}} {\varphi ({\mathbf{x}})H_m^l({\omega _o})\;L({\mathbf{x}},{\omega _o})\;d{\omega _o}} \;d{\mathbf{x}}}. \eqn \]


Similar to density estimation in photon mapping \cite{Jensen1995}, we approximate the spatial integral as a sum of rays $r$ incident on the one-ring neighbourhood of vertex $k$:
\[{L_k}({\omega _o}) = \frac{1}{{{A_k}}}\sum\nolimits_r {{\varphi _k}({\mathbf{x}})f({\mathbf{x}},{-\omega _i},{\omega _o}){\Phi _i}}, \eqn \label{eq:appdxSpatialCoeffs} \]
where ${\Phi _i}$ is the radiant flux transported by a ray arriving at ${\mathbf{x}}$ from direction ${\omega _i}$.

Finally, we approximate the directional integral using Monte Carlo sampling, similar to Krivanek et al.~\shortcite{Krivanek2005}. Consequently, we can write the coeffcients of our radiance cache as a sum over incident rays and local exitant directions:
\[{L_{klm}} = \frac{1}{{N'}}\frac{1}{{{A_k}}}\sum\nolimits_o {\;H_m^l({\omega _o})\;\sum\nolimits_r {{\varphi _k}({\mathbf{x}})f({\mathbf{x}},{-\omega _i},{\omega _o}){\Phi _i}} }, \eqn \label{eq:appdxDirectionalCoeffs} \]
where ${N'}$ is the number of outgoing direction samples ${\omega _o}$ per incident ray.
Considering each incident ray and exitant sample separately yields the update rule stated in Eq.~\eqref{eq:directionalCoeffs}.
Figure~\ref{fg:hsh-coins-fwd} shows a sequence of results for varying resolution of our radiance data structure.

\begin{algorithm}[b]
\caption{Symbolic derivative computation for Eq.~\eqref{eq:posDeriv} }
\label{alg:symbDeriv}
\begin{lstlisting}[]
% MATLAB@\textcolor{darkgreen}{\textregistered}@ Script
x = sym('x',[3,1],'real'); % light position
h = sym('h',[3,1],'real'); % first ray intersection
n = sym('n',[3,1],'real'); % surface normal
syms tinyEps real;         % regularizer

d = x - h;
l = norm(d); % ray length
d = d./l;    % normalized direction

cos_over_r2 = simplify(dot(n, d)/(l*l+tinyEps));

dcos_over_r2dx = gradient( cos_over_r2, x );
cfcn = ccode( simplify( dcos_over_r2dx ) )
% -> copy output to shader code
\end{lstlisting}
\end{algorithm}

\subsection*{Derivatives for Shader Code}

Throughout our implementation we regularly use symbolic math tools to generate code for individual derivative terms.
In Alg.~\ref{alg:symbDeriv} we show this calculation for the $\cos / r^2$-term in Eq.~\eqref{eq:posDeriv} as an instructive example.

\subsection*{Derivative terms for optimization parameters}

In \S5 of the main paper we briefly describe various light source types and their optimization parameters our system currently supports.
Here, we provide more details on how we compute the radiative flux per ray $\Phi_r$ and consequently its derivative with respect to the optimization parameters.
We employ symbolic differentiation and automatic code generation to obtain these derivative terms.
The following derivations only consider a single colour channel, the extension to RGB colours follows the principle described in the main paper.

For a \emph{point light}, each one of $N$ exitant rays covers a solid angle of $4 \pi / N$.
The intensity is assumed uniform over all directions, therefore $\Phi_r = I_e 4 \pi / N$.
The derivative wrt.~the light position is detailed in the main paper.

In the case of a \emph{spot light}, we assume constant intensity $I_e$ in an inner cone, $\theta < \theta_{in}$, where $\theta$ is the angle between the exitant ray direction and the light's world space direction $\mbf{n}$.
In this case the calculation proceeds analogous to the point light case.
Outside of an outer cone, $\theta > \theta_{out}$, the intensity is zero; we only sample rays within this cone, hence the solid angle per ray reduces to $2 \pi (1-cos\theta_{out}) / N$.
Finally, in the case of the spot light's soft edge, $\theta_{in} < \theta \leq \theta_{out}$.
In this region, we assume a quadratic intensity fall-off in terms of $cos \theta$:
\[a(\theta) = {\left( {\frac{{\cos \theta  - \cos {\theta _{out}}}}{{\cos {\theta _{in}} - \cos {\theta _{out}}}}} \right)^2}. \eqn\]
The flux is then $\Phi_r = {I_e 2\pi N^{-1} (1-cos \theta_{out})} \, a(\theta)$, and the derivative wrt.~the light position $\mbf{x}$ follows the same approach as Eq.~\eqref{eq:posDeriv}, but includes a second term for the attenuation function:
\[\frac{{d{\Phi _r}}}{{d{\mathbf{x}}}} = {{\tilde I}_e}\frac{{d(\cos {\omega _1}/{r^2})}}{{d{\mathbf{x}}}} + {I_e}\frac{2\pi}{N} (1 - cos{\theta _{out}})\frac{{da}}{{d{\mathbf{x}}}}. \eqn\]
We compute the last term, $da/d\mbf{x}$, using symbolic differentiation.
Note that $\theta$ is the angle between the light's direction $\mbf{n}$ and the vector $\mbf{x}-\mbf{x}_1$, where $\mbf{x}_1$ is the first ray-surface intersection point, which remains constant as $\mbf{x}$ moves.
Derivatives of $a$, and therefore $\Phi_r$, wrt.~the light's direction $\mbf{n}$ or cone angles $(\theta_{in}, \theta_{out})$ follow analogously.

An \emph{IES light} is a point-shaped light source with a non-uniform intensity profile.
We store the tabulated intensity values specified in an IES data file in a texture, which we interpolate bi-linearly to find $I_e(\theta,\rho)$.
Here $(\theta,\rho)$ are the polar coordinates of the exitant ray direction $\omega_1$ relative to the light's local frame of reference defined by a normal and tangent vector $(\mbf{n},\mbf{t})$.
The derivative wrt.~the light position is then
\[\frac{{d{\Phi _r}}}{{d{\mathbf{x}}}} = {{\tilde I}_e}\frac{{d(\cos {\omega _1}/{r^2})}}{{d{\mathbf{x}}}} + \frac{4\pi}{N} \frac{{\partial {I_e}}}{{\partial (\theta ,\rho )}}\frac{{d(\theta ,\rho )}}{{d{\mathbf{x}}}}, \eqn\]
where we compute both the derivative of the bi-linear interpolation, ${{\partial {I_e}}}/{{\partial (\theta ,\rho )}}$, and the polar coordinates, ${{d(\theta ,\rho )}}/{{d{\mathbf{x}}}}$, using symbolic code generation.
Similarly, the terms required to evaluate Eq.~\eqref{eq:rotVecDeriv} are
\[\frac{{d {\Phi _r}}}{{d {\mathbf{n}}}} = \frac{{4\pi }}{N}\frac{{\partial {I_e}}}{{\partial (\theta ,\rho )}}\frac{{d(\theta ,\rho )}}{{d{\mathbf{n}}}} \eqn \]
for the normal $\mbf{n}$, and analogously for the tangent $\mbf{t}$.

Finally, for \emph{area lights}, we assume that the emission follows a cosine-shaped profile and does not vary across the light's surface.
For convenience, we also scale by the surface area $A$, so that the total emitted power remains constant when changing the light's shape.
Hence the emitted radiance is ${L_e}\cos(\theta)/A$ at every point on the surface, where $\theta$ is the angle between the light's surface normal $\mbf{n}$ and the exitant ray direction $\omega_1$.
For simplicity, we only consider planar, rectangular light shapes in this work.
We again trace $N$ exitant rays, sampled uniformly over the surface and hemisphere, with exitant flux $\Phi_r = L_e \, cos(\theta) /N$.
The flux derivative wrt.~the light's position (i.e.~applying a rigid translation to the light source) is then
\[\frac{{d{\Phi _r}}}{{d{\mathbf{x}}}} = {{\tilde I}_e}\frac{{d(\cos {\omega _1}/{r^2})}}{{d{\mathbf{x}}}} + \frac{{{L_e}}}{N}\frac{{d(\cos \theta )}}{{d{\mathbf{x}}}}. \eqn\]
When optimizing for rotations of area lights, we now must also take into account that the ray origin $\mbf{x}_0$ on the light's surface moves when it is rotated about it's centroid, resulting in
\[\frac{{d{\Phi _r}}}{{d{\mathbf{n}}}} = {{\tilde I}_e}\frac{{d(\cos {\omega _1}/{r^2})}}{{d{\mathbf{n}}}} + \frac{{{L_e}}}{N}\frac{{d(\cos \theta )}}{{d{\mathbf{n}}}}, \eqn\]
and again analogously for the light's tangent vector.
Here, we compute the derivative of $\cos {\omega _1}/{r^2}$ wrt.~the light's orientation by keeping the ray-surface intersection $\mbf{x}_1$ constant, while the ray origin $\mbf{x}_0$ rotates about the light's centroid $\mbf{x}$ such that it's coordinates in the local frame of reference defined by $(\mbf{n},\mbf{t})$ remain unchanged.
Recall that $\omega_1$ is the direction, and $r$ the distance, from $\mbf{x}_0$ to $\mbf{x}_1$.
We again compute the derivative of this transformation by symbolic differentiation and automatic code generation.

\begin{figure}[b]
 \includegraphics[width=\columnwidth]{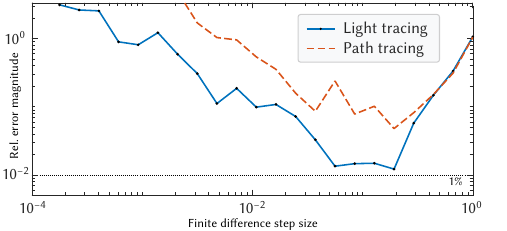}
 \caption{
Finite difference errors of light tracing vs.~path tracing: equal time comparison, where all evaluations start with a new, arbitrary random seed.
 }
\label{fg:fd-pt-nonconst-small}
\end{figure}

\subsection*{Further gradient comparison} 

In Fig.~\ref{fg:fd-pt-nonconst-small} we extend the comparison in  \S\ref{sc:mitsuba} of our adjoint light tracing to a reference differentiable path tracing implementation that also uses our view-independent data structure to deliver the rendered radiance field and evaluate the same objective function.
For a comparison that is indicative of optimization performance, we choose a new arbitrary random seed for each evaluation during finite difference approximation. In this way, the finite difference errors reveal that the path tracing approach produces less accurate gradients (due to sampling noise) in an equal time comparison to our method.
Note that choosing new random seeds is more applicable in an optimization context, where we want sampling noise to average out over the course of the optimization procedure, rather than optimize for a specific noise pattern.

\begin{figure}[t]
 \includegraphics[width=\columnwidth]{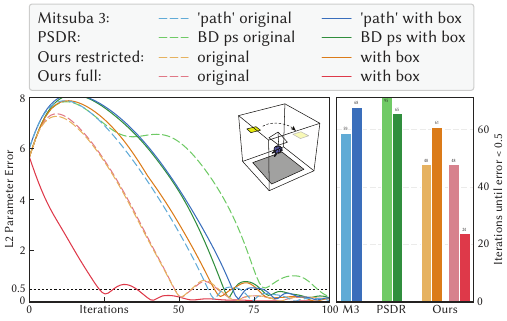}
 \caption{Adding a box around all objects in the test scene of Fig.~\ref{fg:imageCompBunnyOpt}, leads to different convergence behaviour. Using only camera-visible data reduces optimization performance when adding indirect illumination, whereas taking the full radiance field into account improves performance by ca.~$50\%$ in terms of the number of iterations needed to reduce the parameter-space error below $0.5$ (dark red bar and line). We use the same optimization and render settings as in the original test.
 }
\label{fg:baseline-bunny-box}
\end{figure}

\begin{table}[b]
\caption{Summary of all methods and their performance in our comparisons.
(\faCheck:~works, \faRocket:~converges quickly, \scalebox{1.3}{$\mbf{\thicksim}$}:~converges eventually, \rev{}{\faFlash:~available without observable effect,} \faCamera:~image-based, \faCube:~view-independent, \faCamera$/$\faCube:~camera-visible data only)
} \label{tb:methods}
\begin{center}
\begin{tabular}{ l | c c c c }
 Method        & $dO/d\mbf{p}$    & Discont. & Space  & Converges  \\ 
\hline
M3: 'path'        & AD      &      & \faCamera & \faCheck \\
M3: 'ptracer'     & AD      &      & \faCamera &  \\ 
M3: 'prb-reparam' & Adjoint &   \faCheck       & \faCamera & \scalebox{1.3}{$\mbf{\thicksim}$}\\ 
\hline
PSDR: PT ps      & AD      &    \faCheck  & \faCamera & \faCheck \\ 
PSDR: LT ps     & AD      &  \faCheck   & \faCamera & \faCheck \\ 
PSDR: BD ps     & AD      &   \faCheck   & \faCamera & \faCheck \\ 
\hline
Our PT (full)  & Adjoint &      & \faCube & \faRocket\\ 
Our LT (full) & Adjoint &      & \faCube & \faRocket \\ 
Our PT (restr.) & Adjoint &     & \faCamera$/$\faCube & \faCheck \\ 
Our LT (restr.) & Adjoint &    & \faCamera$/$\faCube & \faCheck \\ 
Our LT (na\"ive) & Direct &      & \faCube &  \\ 
\end{tabular}
\end{center}
\end{table}

\subsection*{Extended baseline comparison}
In our baseline test case (Fig. \ref{fg:imageCompBunnyOpt}) we intentionally avoid most global influences caused by indirect illumination.
As expected, our novel view-independent approach performs comparable to state-of-the-art image-based methods, regardless of whether we restrict our method to camera-visible surfaces, or evaluate the entire scene.
In Fig.~\ref{fg:baseline-bunny-box}, we now extend the test scene by adding a box around all other objects, which increases the contribution of indirect illumination to the bunny and the background plane. Mitsuba's path tracer (without discontinuity handling), as well as our method when using restricted data for comparison, now converge slightly slower than in the original case (without the box), showing that the optimization problem becomes more challenging due to indirect illumination.
PSDR (including discontinuity handling) now matches the convergence of these methods (having been worse before), showing that the additional indirect illumination reduces the influence of the discontinuous gradient contributions.
Finally, using our method and the full radiance data (including the box) as a target, improves the convergence rate by roughly $50\%$, demonstrating the advantage of a view-independent approach.
When completing the entire optimization ($100$~iterations), our method is about $3 \times$ faster than Mitsuba's path tracer on this scene.
Conversely, for image-based methods we would need to introduce additional cameras that cover the new surfaces of the box to achieve a similar effect. Table~\ref{tb:methods} gives an overview of all methods used and summarizes their features.

\subsection*{Gradient contribution visualization}

\begin{figure}[b]
\def\svgwidth{\columnwidth}
\begingroup%
  \makeatletter%
  \providecommand\color[2][]{%
    \errmessage{(Inkscape) Color is used for the text in Inkscape, but the package 'color.sty' is not loaded}%
    \renewcommand\color[2][]{}%
  }%
  \providecommand\transparent[1]{%
    \errmessage{(Inkscape) Transparency is used (non-zero) for the text in Inkscape, but the package 'transparent.sty' is not loaded}%
    \renewcommand\transparent[1]{}%
  }%
  \providecommand\rotatebox[2]{#2}%
  \newcommand*\fsize{\dimexpr\f@size pt\relax}%
  \newcommand*\lineheight[1]{\fontsize{\fsize}{#1\fsize}\selectfont}%
  \ifx\svgwidth\undefined%
    \setlength{\unitlength}{243.77953837bp}%
    \ifx\svgscale\undefined%
      \relax%
    \else%
      \setlength{\unitlength}{\unitlength * \real{\svgscale}}%
    \fi%
  \else%
    \setlength{\unitlength}{\svgwidth}%
  \fi%
  \global\let\svgwidth\undefined%
  \global\let\svgscale\undefined%
  \makeatother%
  \begin{picture}(1,1.10204082)%
    \lineheight{1}%
    \setlength\tabcolsep{0pt}%
    \put(0,0){\includegraphics[width=\unitlength,page=1]{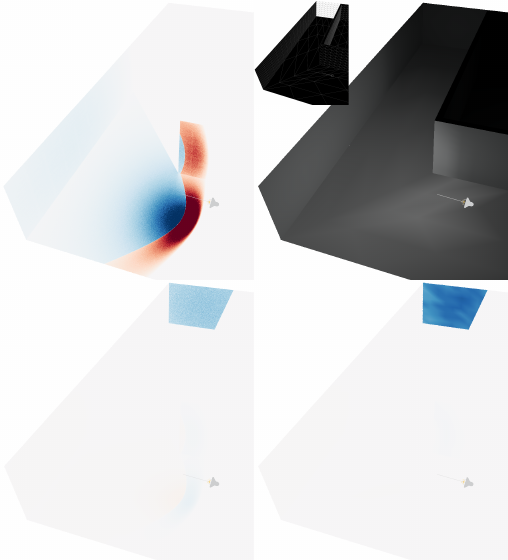}}%
    \put(0.15440202,1.04938545){\makebox(0,0)[lt]{\lineheight{1.25}\smash{\begin{tabular}[t]{l}(a)\end{tabular}}}}%
    \put(0.69231413,1.04938545){\makebox(0,0)[lt]{\lineheight{1.25}\smash{\begin{tabular}[t]{l}(b)\end{tabular}}}}%
    \put(0.15440202,0.48030137){\makebox(0,0)[lt]{\lineheight{1.25}\smash{\begin{tabular}[t]{l}(c)\end{tabular}}}}%
    \put(0.69231413,0.48030137){\makebox(0,0)[lt]{\lineheight{1.25}\smash{\begin{tabular}[t]{l}(d)\end{tabular}}}}%
    \put(0.56348223 ,0.36680303){\makebox(0,0)[lt]{\lineheight{1.25}\smash{\begin{tabular}[t]{l}$(\partial O / \partial L) (\Delta L / \Delta z)$\end{tabular}}}}%
    \put(0.04455442,0.36680303){\makebox(0,0)[lt]{\lineheight{1.25}\smash{\begin{tabular}[t]{l}$(\partial O / \partial L) (dL/dz)$\end{tabular}}}}%
    \put(0.04455442,0.93105301){\makebox(0,0)[lt]{\lineheight{1.25}\smash{\begin{tabular}[t]{l}$(\partial O / \partial \Phi_r) (d\Phi_r /dz)$\end{tabular}}}}%
  \end{picture}%
\endgroup%

\caption{
Our gradient contribution visualization under indirect illumination, showing derivatives of the objective wrt.~the $z$ coordinate (left-to-right in this view) of a spot light.
Our method (a) visualizes contributions of adjoint states at the first ray-surface intersection of light paths reaching the target (b, inset) after some indirect bounces.
Note that paths traversing points in the outer cone of the spot light (where the light's intensity is attenuated) would increase in brightness if the light were to move to the right, while most other surfaces prefer the light moving left.
Direct (non-adjoint) differentiation (c), or finite difference approximation (d) instead visualize indirect gradient contributions at the target surface itself.
Image (b) shows the global illumination solution.
}
\label{fg:grad-vis-indirect}
\end{figure}

In this section, we show extended comparisons for our gradient contribution visualization introduced in \S\ref{sc:gradVis}.
We first expand Fig.~\ref{fg:grad-vis-dragon} with additional comparisons showing all three light position coordinates in Fig.~\ref{fg:grad_vis_full}.
We again show results for finite differencing ($\Delta L / \Delta \mbf{p}$), direct (i.e.~non-adjoint) differentiation ($dL/d\mbf{p})$, and our adjoint state ($\partial O/\partial \Phi_r$).

We then show an additional visualization example, Fig.~\ref{fg:grad-vis-indirect} for an intermediate lighting configuration that occurs during the optimization shown in Fig.~\ref{fg:indirectOpt}.
At this point, the spot light has not navigated around the final corner in the labyrinth, so the target is still only illuminated by indirect light.
In this situation, the difference between our adjoint approach, and a direct differentiation method (akin to forward-mode autodiff) is most clearly visible.
Direct differentiation produces a similar result as a finite difference approximation, visualizing contributions to the objective function gradient at the target surface (after indirect light arrives there).
Our approach, on the other hand, transports information about objective function derivatives backwards along light paths and visualizes the resulting gradient contribution at the first ray-surface intersection along these paths.
In this way, we can observe how (indirect) light paths passing through parts of the surface near the light source contribute to the objective gradient at the target surface.
Note that the objective gradient (which is integrated over all such contributions on surfaces) still pulls the light left towards the target.

\begin{figure*}[p]
 \includegraphics[width=0.9\textwidth]{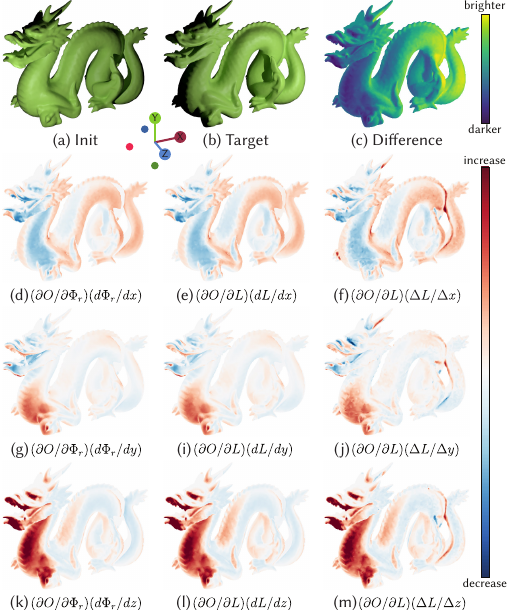}
 \caption{
Extended version of Fig.~\ref{fg:grad-vis-dragon}, our gradient visualization on the Stanford Dragon.
The second to fourth rows show objective function gradient contributions for each position coordinate of the light source. Columns show our adjoint gradients (left), direct differentiation (centre), and finite difference approximation (right). Image subtitles indicate the calculation; BRDF derivatives are taken into account but not explicitly stated here for brevity.
 }
\label{fg:grad_vis_full}
\end{figure*}

\clearpage

\markupsfalse 

\ifmarkups

\clearpage

\subsection*{ *** WORK IN PROGRESS ***}
\todo{ These work in progress parts will eventually end up in either the main paper, the appendix, or a supplement ... let's first collect everything here and organize them later. 
}
---

\begin{figure}[h]
 \includegraphics[width=\columnwidth]{FD-naive_simpleOffice-finiteDiff-comp-axis-all.pdf}
 \caption{
Finite difference validation: our gradients compared to a naive implementation that directly computes derivatives of the ray-surface intersections when moving a point light, using 38m light rays.
 }
\label{fg:fd-naive}
\end{figure}

\begin{figure}
 \includegraphics[width=\columnwidth]{FD-naive_lowray_nobounce_simpleOffice-finiteDiff-comp.pdf}
 \caption{
Finite difference validation: our gradients compared to a naive implementation that directly computes derivatives of the ray-surface intersections when moving a point light, using 16k light rays.
 }
\label{fg:fd-naive-lowray}
\end{figure}

Figure~\ref{fg:fd-naive} shows a comparison of finite difference errors over a range of FD step sizes $h$.
Here we compute the gradient of the optimization objective wrt.~the light's position for the initial state as shown in Fig.~\ref{fg:simpleOffice-opt-comp}a
This test shows that a naive implementation, which directly differentiates the motion of the ray-surface intersection point when changing the position of a light source, does not correctly account for the derivative of the geometric term $cos_{\text{in}} / r^2$, as evidenced by the systematic error seen mostly in the $y$ component of the gradient.
Our gradient calculation, on the other hand, explicitly differentiates this geometric term and leads to much smaller errors for a wide range of step sizes.
\todo{ also refer to the optimization behaviour ... replace the old example with a test on the simple office scene? }
Furthermore, when using a low number of rays ($16$k instead of $38$m) in Fig.~\ref{fg:fd-naive-lowray}, we can see that the naive gradient calculation is correct for very small step sizes (so long as no ray-surface intersection crosses from one triangle to a neighbouring one), but for larger step sizes---which are important from an optimization point of view---our method delivers better results even when using a low number of samples (where rounding errors are less pronounced).

---


\begin{figure}
 \includegraphics[width=\columnwidth]{FD-PT_simpleOffice-finiteDiff-comp.pdf}
 \caption{
Finite difference errors of light tracing vs.~path tracing, both using our view-independent data structure and 38m primary samples (leaving the light source or the scene geometry respectively). All evaluations follow the same pseudo-random sequence.
 }
\label{fg:fd-pt}
\end{figure}

\begin{figure}
 \includegraphics[width=\columnwidth]{FD-PT_nonconst_simpleOffice-finiteDiff-comp.pdf}
 \caption{
Finite difference errors of light tracing vs.~path tracing: equal time comparison, where all evaluations start with a new, arbitrary random seed.
 }
\label{fg:fd-pt-nonconst}
\end{figure}

In Fig.~\ref{fg:fd-pt} and \ref{fg:fd-pt-nonconst} we compare our adjoint light tracing to an implementation of the path-replay backpropagation method that also uses our view-independent data structure to deliver the rendered radiance field and evaluate the same objective function.
For very small step sizes, rounding errors are smaller for the path-traced version, as the forward rendering samples the light source via next event estimation, therefore also explicitly evaluating the geometric term $cos_{\text{in}} / r^2$.
For larger step sizes that are more indicative of optimization performance, the two methods are evenly matched when each evaluation during the FD-approximation follows the same pseudo-random sequence that was also used to estimate the adjoint gradient.
However, when using an arbitrary random seed in each evaluation, Fig.~\ref{fg:fd-pt-nonconst} the path tracing approach produces more gradient errors (due to sampling noise) in an equal time comparison.
Note that setting new random seeds is also more applicable in an optimization context, where we want sampling noise to average out over the course of the optimization procedure, rather than optimize for a specific noise pattern.

\fi
\fi

\end{document}